\documentclass[a4paper,fleqn,usenatbib]{mnras}

\usepackage{newtxtext,newtxmath}
\usepackage[T1]{fontenc}
\usepackage{ae,aecompl}

\usepackage[utf8]{inputenc}
\usepackage{natbib}
\usepackage{threeparttablex} % tablenotes
\usepackage{booktabs} % cmidrule
\usepackage{longtable}
\usepackage{graphicx}
\usepackage{rotating}
\usepackage{amssymb, amsmath, amsbsy}
\usepackage{url} 
\newcommand{\rrab}{RR{\sl ab}}
\newcommand{\rrc}{RR{\sl c}}
\newcommand{\rrd}{RR{\sl d}}
\newcommand{\rrcd}{RR{\sl cd}}

\title[Variable Stars in Crater II]{A DECam View of the Diffuse Dwarf Galaxy Crater II: Variable Stars}

\author[A. K. Vivas et al.]{A. Katherina Vivas,$^{1}$\thanks{Contact e-mail: \href{mailto:kvivas@ctio.noao.edu}{kvivas@ctio.noao.edu}}
Alistair R. Walker,$^{1}$
Clara E. Mart{\'\i}nez-V{\'a}zquez,$^{1}$
\newauthor
Matteo Monelli,$^{2,3}$
Giuseppe Bono,$^{4}$
Antonio Dorta,$^{2,3}$
David L. Nidever,$^{5,6}$
\newauthor
Giuliana Fiorentino,$^{7}$
Carme Gallart,$^{2,3}$
Gloria Andreuzzi,$^{8,9}$
Vittorio F. Braga$^{10,11}$
\newauthor
Massimo Dall'Ora,$^{12}$
Knut Olsen,$^{6}$
Peter B. Stetson,$^{13}$
\\
$^{1}$Cerro Tololo Inter-American Observatory, NSF's National Optical-Infrared Astronomy Research Laboratory, Casilla 603,
La Serena, Chile\\
$^{2}$Instituto de Astrof{\'\i}sica de Canarias, Calle V{\'\i}a L\'actea, E-38205 La Laguna, Tenerife, Spain\\
$^{3}$Universidad de La Laguna, Dpto. Astrof{\'\i}sica, E-38206 La Laguna, Tenerife, Spain\\
$^{4}$Departimento di Fisica, Universitá di Roma Tor Vergata, via della Ricerca Scientifica 1, I-00133 Rome, Italy\\
$^{5}$Department of Physics, Montana State University, P.O. Box 173840, Bozeman, MT 59717-3840, USA\\
$^{6}$NSF's National Optical-Infrared Astronomy Research Laboratory, 950 North Cherry Avenue, Tucson, AZ 85719, USA\\
$^{7}$INAF - Osservatorio Astronomica di Bologna, via Ranzani 1, I-40127 Bologna, Italy\\
$^{8}$ INAF - Fundaci\'on Galileo Galilei, Rambla Jos\'e Ana Fernandez P\'erez 7, 38712, Bre{\~n}a Baja, Spain \\
$^{9}$ INAF - Osservatorio Astronomico di Roma, via Frascati 33, 00040, Monte Porzio Catone (Roma), Italy \\
$^{10}$ Instituto Milenio de Astrof{\'\i}ısica, Santiago, Chile \\
$^{11}$ Departamento de F{\'\i}sica, Facultad de Ciencias Exactas, Universidad Andr\'es Bello, Fern\'andez Concha 700, Las Condes, Santiago, Chile \\
$^{12}$INAF - Osservatorio Astronomica di Capodimonte, salita Moiariello 16, I-80131 Napoli, Italy\\
$^{13}$Dominion Astrophysical Observatory, Herzberg Institute of Astrophysics, National Research Council, Victoria, British Columbia V9E 2E7, Canada\\
}

\date{Accepted XXX. Received YYY; in original form ZZZ}

\pubyear{2019}

\begin{document}
\label{firstpage}
\pagerange{\pageref{firstpage}--\pageref{lastpage}}
\maketitle

\begin{abstract}
Time series observations of a single dithered field centered on the diffuse dwarf satellite galaxy Crater II were obtained with the Dark Energy Camera (DECam) at the 4m Blanco Telescope at Cerro Tololo Inter-American Observatory, Chile, uniformly covering up to two half-light radii. Analysis of the $g$ and $i$ time series results in the identification and characterization of 130 periodic variable stars, including 98 RR Lyrae stars, 7 anomalous Cepheids, and 1 SX Phoenicis star belonging to the Crater II population, and 24 foreground variables of different types. Using the large number of ab-type RR Lyrae stars present in the galaxy, we obtained a distance modulus to Crater II of $(m-M)_0=20.333\pm 0.004$ (stat) $\pm 0.07$ (sys). The distribution of the RR Lyrae stars suggests an elliptical shape for Crater II, with an ellipticity of 0.24 and a position angle of $153\degr$. From the RR Lyrae stars we infer a small metallicity dispersion for the old population of Crater II of only 0.17 dex. There are hints that the most metal-poor stars in that narrow distribution have a wider distribution across the galaxy, while the slightly more metal rich part of the population is more centrally concentrated. Given the features in the color-magnitude diagram of Crater II, the anomalous Cepheids in this galaxy must have formed through a binary evolution channel of an old population.
\end{abstract}


\section{Introduction} \label{sec:intro}

Crater II is a fascinating satellite galaxy discovered by \citet{torrealba16} in the VST ATLAS Survey at $\sim 115$ kpc from the Sun. With a half-light radius ($r_h$) of $31\farcm 2$, it is one of the largest satellite galaxies of the Milky Way, only surpassed by the Magellanic Clouds, Sagittarius and the recently discovered Antlia II galaxy \citep{torrealba19}. Together with Antlia II, these two dwarf galaxies extend the population of satellite galaxies to a low-luminosity, large size regime not known before among the satellites of the Milky Way. They lie in the frontier between classical satellites and ultra-faint dwarf (UFD) galaxies. Crater II may have suffered, and may still be suffering, heavy mass loss and strong tidal evolution as suggested by its orbit around the Milky Way \citep{sanders18,fritz18,fu19}, which brings it as close as 33 kpc from the Galactic center at perihelion every 2.1 Gyr. Observational evidence of tidal debris has yet to be discovered, however.

Hence, Crater II presents an opportunity to study the stellar population(s) of a galaxy under strong tidal interaction with the Milky Way. We carried out an intensive observing campaign on Crater II in order to study its stellar content in detail. Our approach is twofold: high cadence time series observations are used to characterize its variable star population, while the stacking of our multi-epoch observations of the galaxy produce deep  color-magnitude diagrams (CMD). This paper presents the former approach while the latter is presented in a companion article \citep{walker19}.

Variable stars have a long tradition of being tracers of different stellar populations. The most well known of the pulsating variable stars in dwarf galaxies are RR Lyrae stars, which unequivocally trace an old population \citep[$>10$ Gyr,][]{smith95}. RR Lyrae stars have been found in almost all of the satellites of the Milky Way \citep[recent compilation in][]{martinez19}, even in low luminosity systems such as Segue I \citep{simon11}. This is a clear indication that an old population is ubiquitous among dwarf galaxies. On the other hand, anomalous Cepheids, which are pulsating stars above the horizontal branch, may trace an intermediate age population. Their progenitors are likely massive (1-2 $M_\odot$) stars that have evolved off the main sequence \citep{fiorentino12a}. These stars have been commonly found in the classical satellites of the Milky Way but are rare in globular clusters \citep{clement01}, which contain exclusively old populations. Since they have been observed also in dwarf spheroidal (dSph) galaxies with predominantly old stellar populations like Draco and Ursa Minor \citep{nemec88,kinemuchi08}, other formation channels are needed for these stars. Binary evolution can also bring stars to the region of the instability strip where anomalous Cepheids live \citep{gautschy17}. Thus, old progenitors are also possible. Dwarf Cepheid stars (a.k.a. SX Phoenicis and/or $\delta$ Scuti stars) are also pulsating stars in the instability strip but they are found below the horizontal branch \citep{breger00}. Their very short periods (just a few hours) and small amplitudes make them very hard to detect, especially in distant systems. They may trace either intermediate-age populations (massive main sequence stars in the instability strip) or old populations (pulsating blue stragglers).  Finally, Classical Cepheids indicate the presence of young populations (0.03-0.7 Gyrs). They are rare in dSph galaxies, with Leo I being an exception \citep{fiorentino12b}.

Soon after the discovery of Crater II by \citet{torrealba16}, two different teams searched for RR Lyrae stars in Crater II. \citet{joo18} searched for RR Lyrae stars using the Korea Microlensing Telescope Network (KMTNet) 1.6 m telescope, at Cerro Tololo Inter-American Observatory, Chile, in a region of $3\degr \times 3\degr$. They found 96 RR Lyrae stars, of which 97\% were located inside a radius of 1\degr from the center of Crater II. Their search for variables, however, was limited to the horizontal branch region. Thus, no other types of pulsating stars were identified in Crater II. 
Then, \citet{monelli18} independently searched for variable stars using the Issac Newton Telescope (2.54m), at Observatorio del Roque de Los Muchachos, Canary Islands, Spain. The spatial coverage of their search, however, included only the central part of Crater II (a region of $0.44$ deg$^2$).  Within this small region, no other type of variables besides RR Lyrae stars were found. In this work, we make a new search for variable stars in Crater II. The advantages of our approach with respect to these previous works are: {\sl (i)} the use of a 4m telescope allows to obtain excellent SNR at the level of the horizontal branch of Crater II. This allows us to be sensitive to low amplitude variables as well as to variables below the horizontal branch; {\sl (ii)} the large field of view (FoV) of DECam allows us to cover the galaxy out to $\sim 2r_h$ in a single pointing. This is a coverage area similar to the one in \citet{joo18}, but significantly larger than the one in \citet{monelli18}; {\sl (iii)} continuous coverage during 3 consecutive nights allows us to obtain excellent phase coverage and recover periods with no impact from aliasing; and {\sl (iv)} we search for periodic variable stars of all types in our data, not only RR Lyrae stars. Our data, however, are not sensitive to long period variables ($P>2$ d).

This paper is structured as follows. We describe the observations in \S~\ref{sec:observations}, and data processing, photometry and identification of variable stars in \S~\ref{sec:photometry}. In section \S~\ref{sec:rrl}, we discuss the characteristics of the RR Lyrae stars identified in this work. We compare with the previous findings of \citet{joo18} and \citet{monelli18}, discuss the Bailey Diagram, derive a robust distance to Crater II, and examine the spatial distribution and radial trends of the RR Lyrae stars within the galaxy. The population of anomalous Cepheids in Crater II is discussed in \S~\ref{sec:AC}. Other variable stars in the field, including a single SX Phoenicis star in Crater II, are described in \S~\ref{sec:othervar}. Finally, conclusions are summarized in \S~\ref{sec:discussion}.

\section{Observations} \label{sec:observations}

\begin{figure*}
\includegraphics[width=1.0\textwidth]{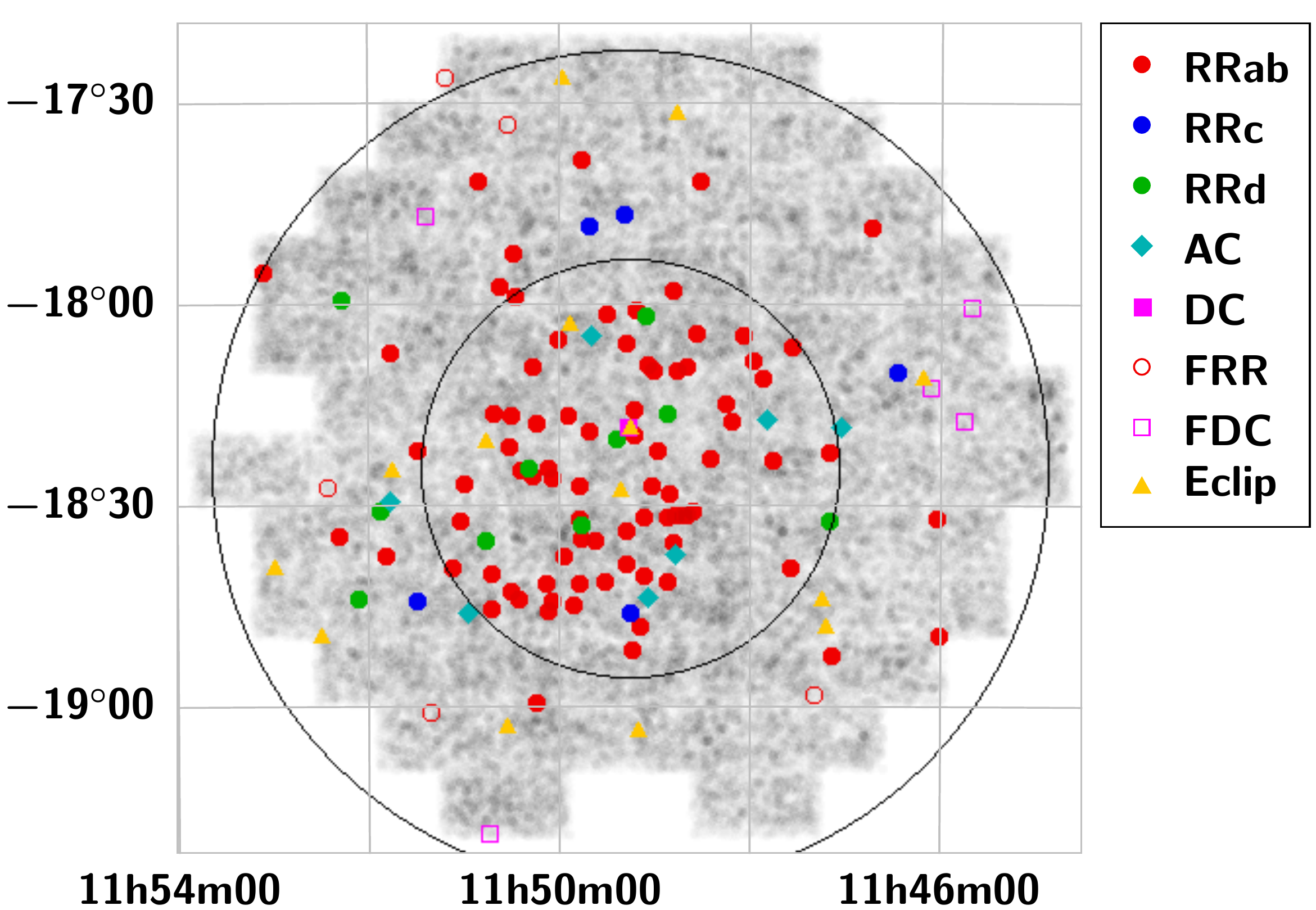}
\caption{Spatial distribution in equatorial coordinates of the 130 periodic variable stars classified in this work. The legend identifies the types of variables (RRab $=$ type ab RR Lyrae star; RRc $=$ type c RR Lyrae star; RRd $=$ type d RR Lyrae star; AC $=$ anomalous Cepheid;  DC $=$ dwarf Cepheid star; FDC $=$ Field dwarf Cepheid star; FRR $=$ Field RR Lyrae star; Eclip $=$ eclipsing binary star). Black circles have radii of $0\fdg 52$ and $1\fdg 04$ ($1\, r_h$ and $2\, r_h$, respectively). The grey background, which shows the footprint of our survey, is a density map made with all stars in our catalog. We note that the CCD gaps are well covered, thanks to the dithering pattern applied to the observations.}
\label{fig:Sky}
\end{figure*}

Observations were carried out using the Dark Energy Camera \citep[DECam,][]{flaugher15} on the 4m Blanco Telescope at Cerro Tololo Inter-American Observatory, Chile.
With a field of view of 3 sq degrees, a single DECam field allows us to cover an area extending to about twice the $r_h$ of the galaxy (Figure~\ref{fig:Sky}), which was measured to be $31\farcm 2$
by \citet{torrealba16}. The observations were carried out during 3 consecutive nights, from March 19 to March 21, 2017, with a few additional exposures taken two weeks later, on the night of April 4, 2017. Crater II culminated near the middle of the night during the March nights, allowing continuous observation of the galaxy for about 9 hours during each of the three nights. Our strategy consisted of observing a single field in filters $g$ and $i$, with the galaxy centered on CCD N4 (one of the central CCDs in the camera). Exposure times were 180s in both filters. In total, we gathered 160 and 158 observations in $g$ and $i$, respectively. The pointing coordinates were $(\alpha,\delta)=(177\fdg 31,-18\fdg 3308)$. Consecutive observations were dithered in a "Center $+$ Rectangle" pattern, with offsets of $45\arcsec$ and $60\arcsec$ in right ascension and declination, respectively. Consequently, the resulting spatial coverage was quite homogeneous, although the average number of exposures in the gap regions drops to $\sim 95$ in each band.  

The seeing of the observations varied between $0\farcs 8$ and $2\farcs 2$ with median values of $1\farcs 2$ in $g$ and $1\farcs 0$ in $i$. Although during the observing nights Crater II spent most of the time with a low airmass (near $1.0$), we pushed the time baseline to be as large as possible in order to have a better coverage of the light curves; thus, we started/ended the observations each night with an airmass near $2.2$. The Moon had an age of $\sim 8$ days during the main observing run. Some $u$-band observations were taken during the time of the night with the Moon down. Those observations were not used for the time series analysis. Conditions were clear and photometric on the last two of the March nights, with scattered clouds during the first night.

On photometric nights, standard star fields in the SDSS footprint were observed at different airmasses.

Image data are available through the NOAO Science archive\footnote{\url{http://archive.noao.edu}}.

\section{Data Reduction and Photometry} \label{sec:photometry}

\begin{figure}
\centering
\includegraphics[width=1.0\columnwidth]{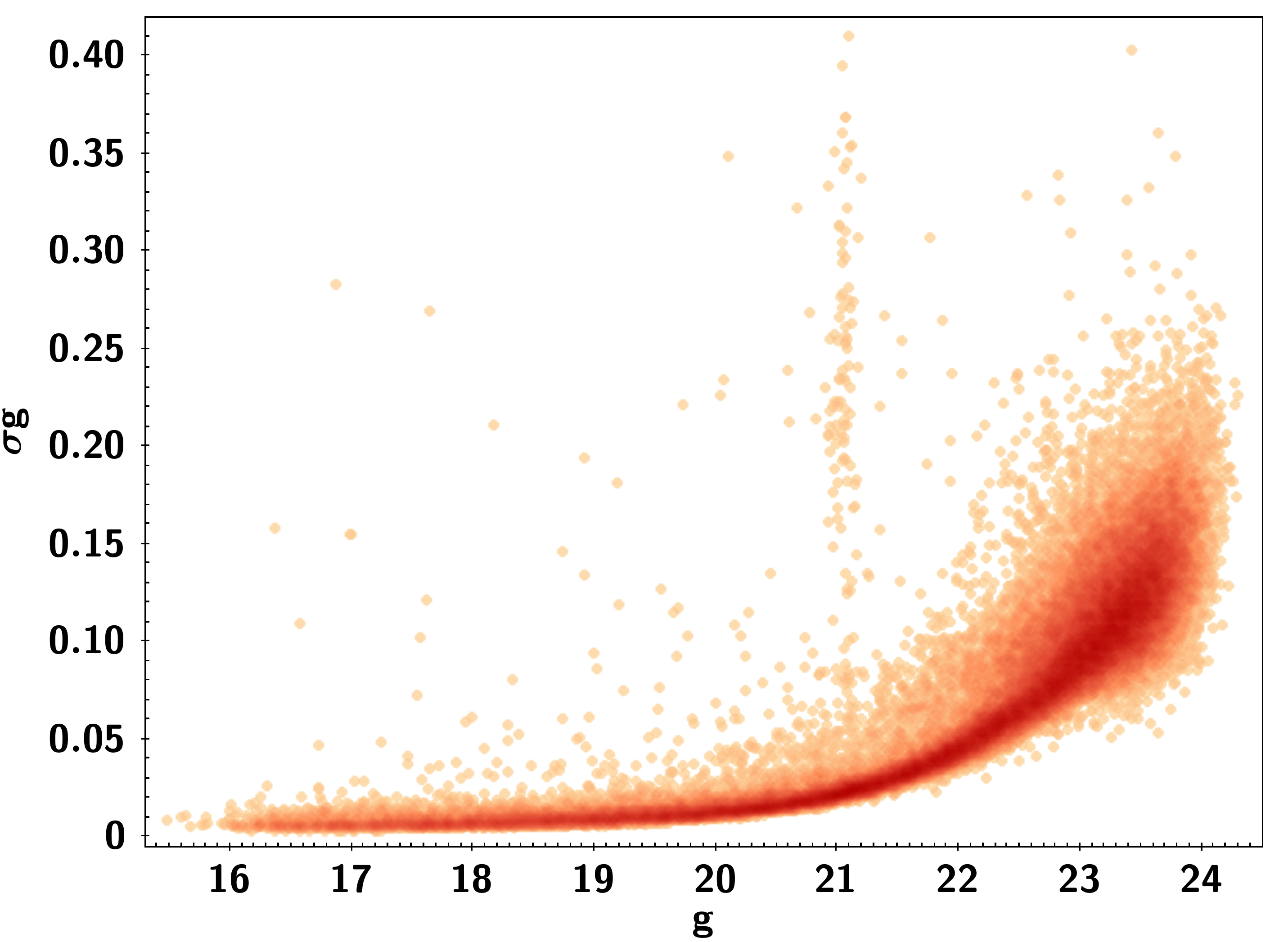}
\includegraphics[width=1.0\columnwidth]{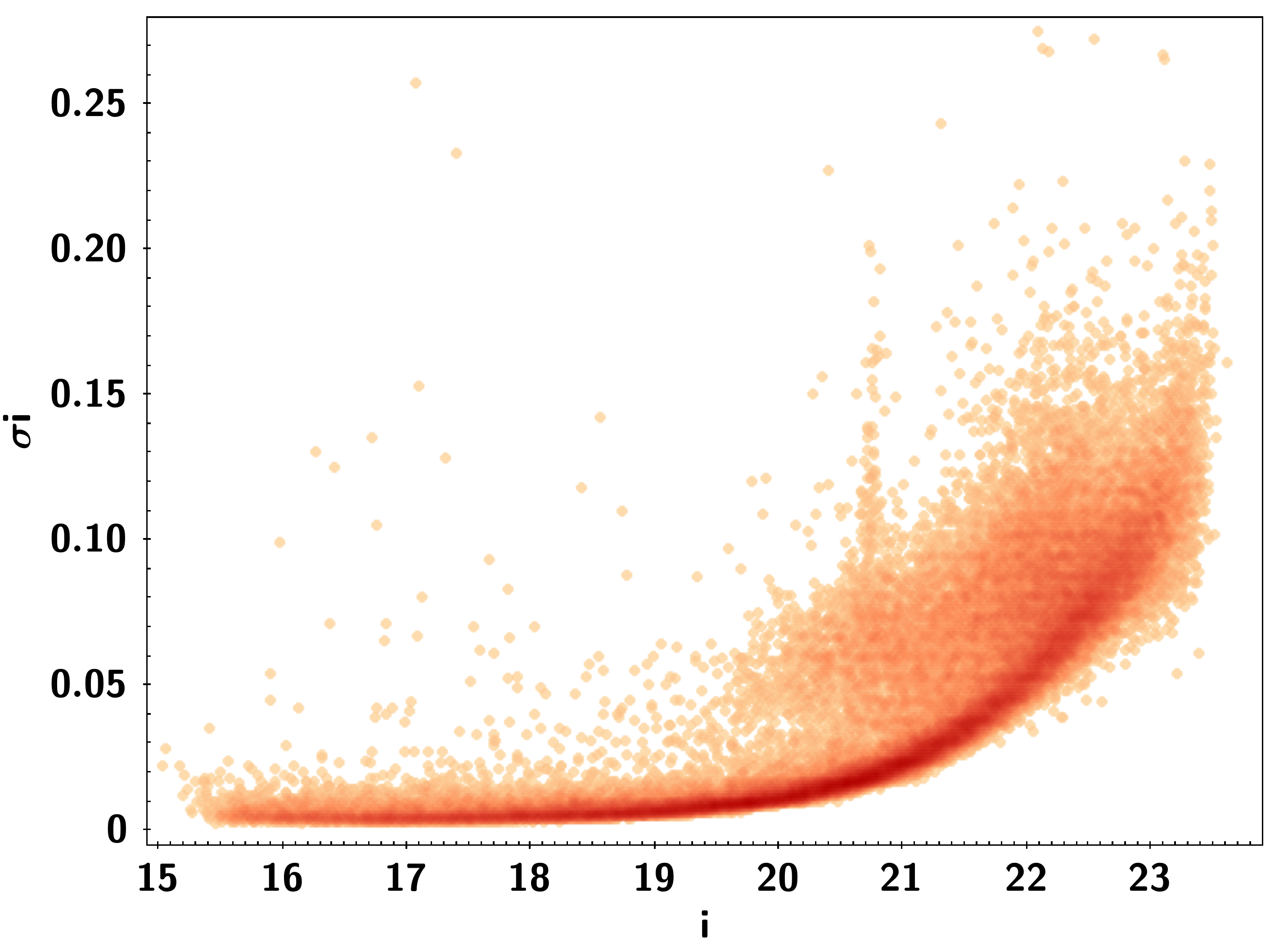}
\caption{Standard deviation of the magnitude distribution of each star in $g$ (Top) and $i$ (Bottom) as a function of mean magnitude.}
\label{fig:error}
\end{figure}

The procedure for image reduction and processing closely follows the one used for the Survey of the MAgellanic Stellar History (SMASH), which is described in detail in \citet{nidever17}. In summary, data were processed by the NOAO Community Pipeline \citep{valdes14} and then, point-spread function (PSF) photometry was extracted using the PHOTRED pipeline \citet{nidever17}, which is based on DAOPHOT \citep{stetson87}. Because the goal of this work was to construct time-series measurements, and because the variable stars of interest were not particularly faint, we skipped some steps in PHOTRED which were not relevant to this work. In particular, we used the photometry given by ALLSTAR directly, and we did not attempt to run ALLFRAME on the data to make detections on the (deeper) stacked data.  Deep photometry from this dataset based on stacked images is presented in a companion paper \citep{walker19}.

Two DECam CCDs, namely \#31 and \#61 (N30 and S7), were not used in this work. N30 does not produce signal while one of the amplifiers in S7 shows unstable gain which makes it unsuitable for precision photometry. 

Photometric calibration was made using the PHOTRED package, with the standard stars observed each night. Zero points and color ($g-i$) terms were determined for each CCD, while unique  atmospheric extinction terms were obtained using all standard stars in all CCDs. Similar to SMASH, the calibration obtained this way is on the SDSS photometric system.

The catalog was cleaned out of  extended objects by discarding those with DAOPHOT parameters $\vert {\rm SHARP} \vert \geq 1.0$ and $\chi >3.0$. Our $g,i$ working catalog contains multi-epoch photometry of 366,237 objects in the field of Crater II.

In order to take into account the non-photometric conditions suffered during one of the nights, we selected exposures in both bands which were obtained with the best sky conditions in order to serve as zero point references for the rest of the observations.  The chosen observations in $g$ and $i$, exposures 632593 and 632594, were taken on March 20/21, 2019 (a photometric night), at an airmass of 1.02 and under excellent seeing conditions ($0\farcs 86$ and $0\farcs 80$ in $g$ and $i$, respectively). We determined zero point differences between all exposures in our survey with respect to the chosen references by obtaining a clipped mean of the difference in magnitude of stars in the range $16 \leq g \leq 21$, within which stars have small photometric errors. This was done independently for every CCD.  Typically 100-250 stars were used for each CCD to determine the zero point. A few exposures for which the standard deviation of the zero points was large ($>0.03$ mags) were eliminated. Finally, we kept only objects that had $\geq15$ measurements in both bands which reduced the size of the catalog to 45,623 stars.

The standard deviation of the magnitude distribution for each star ($\sigma_{\rm star}$) as a function of mean magnitude ($m$) is shown in Figure~\ref{fig:error}. As expected, most of the stars are non-variable and occupy a locus that has increasing $\sigma_{\rm star}$ as a function of $m$. This locus indicates that the photometric errors of our ensemble photometry increase to 0.1 mag at $g=23.2$ and $i=23.0$. Stars above the main locus are variable star candidates. In particular, there is clearly an excess of variable star candidates at $g\sim 21$, which is located at the expected magnitude of RR Lyrae stars in the Crater II galaxy.  Formally, we flagged stars as variable candidates if $\sigma_{\rm star} > \sigma(m) +  3 \times {\rm std}(m)$, where $\sigma(m)$, ${\rm std}(m)$ are the mean value and rms of $\sigma_{\rm star}$ for stars in the non-variable locus in bins of 0.20 magnitudes.

\subsection{Extinction} \label{sec:ext}

\begin{figure}
\includegraphics[width=1.0\columnwidth]{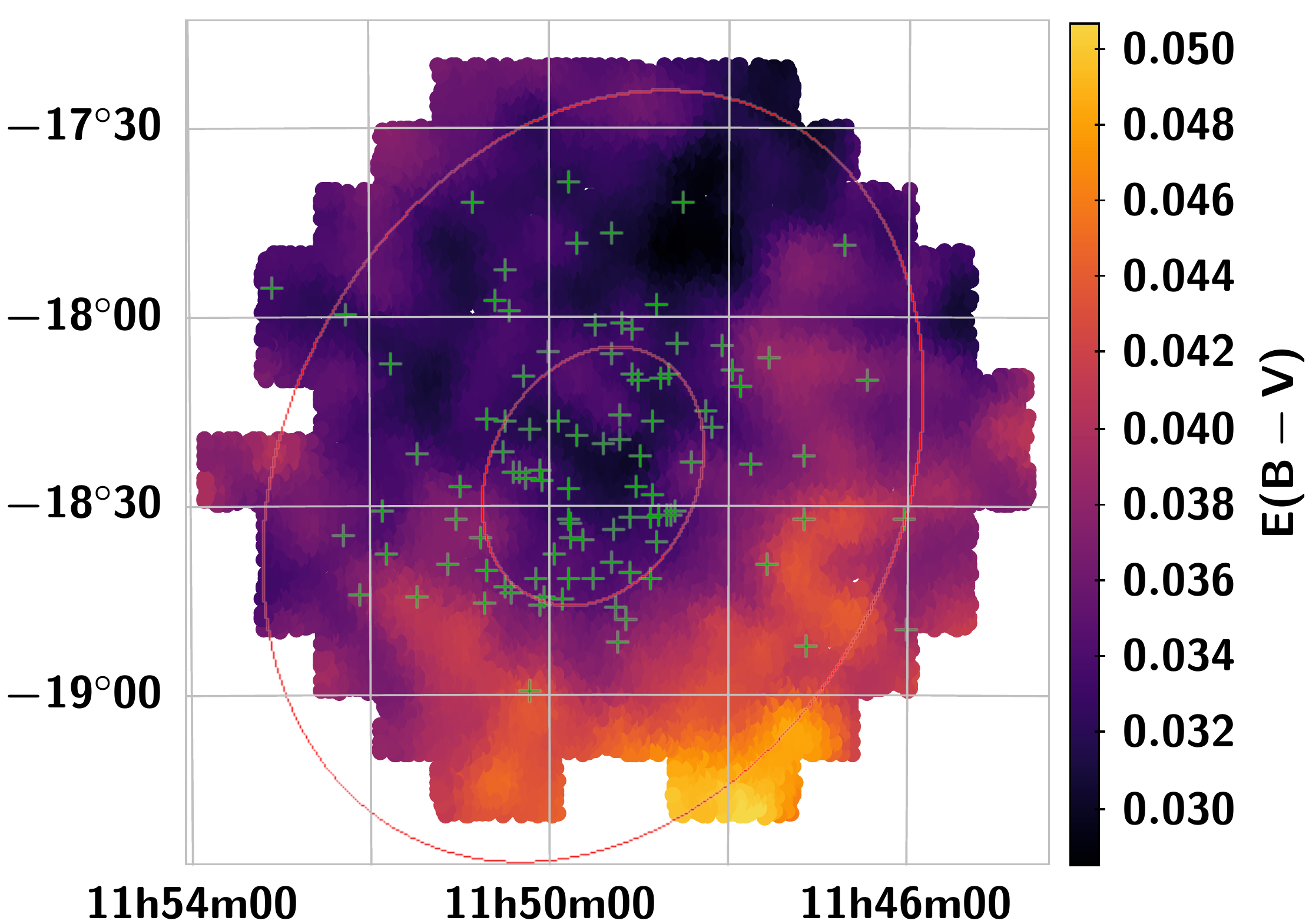}
\caption{Reddening map, E(B-V)$_{\rm SFD}$, in the Crater II region from \citet{schlegel98}. Green asterisks mark the location of the RR Lyrae stars identified in this work. Red lines represent the best fitted $1\sigma$ and $3\sigma$ ellipses to the population of RR Lyrae stars (see \S~\ref{sec:sp}). To obtain the extinction $A_g$ and $A_i$ we used the coefficients derived by \citet{schlafly11}.} 
\label{fig:ext}
\end{figure}

Being located at high Galactic latitude ($b\sim42\degr$), the extinction toward Crater II is quite low. Figure~\ref{fig:ext} shows the reddening map constructed using the color excess, E(B-V)$_{\rm SFD}$, from individual stars in the dust maps\footnote{\url{https://irsa.ipac.caltech.edu/applications/DUST/}} of \citet{schlegel98}. Extinctions in $g$ and $i$ were calculated for each individual star using the re-calibration of the dust maps by \citet{schlafly11}: $A_g = 3.303\times E(B-V)_{\rm SFD}$ and $A_i=1.698\times E(B-V)_{\rm SFD}$. The color excess E(B-V)$_{\rm SFD}$ varies in the observed region from 0.029 to 0.051 mag, with a mean value of 0.036 mag.

\subsection{Variable Star Classification} \label{sec:variables}

\begin{figure*}
\includegraphics[width=0.8\textwidth]{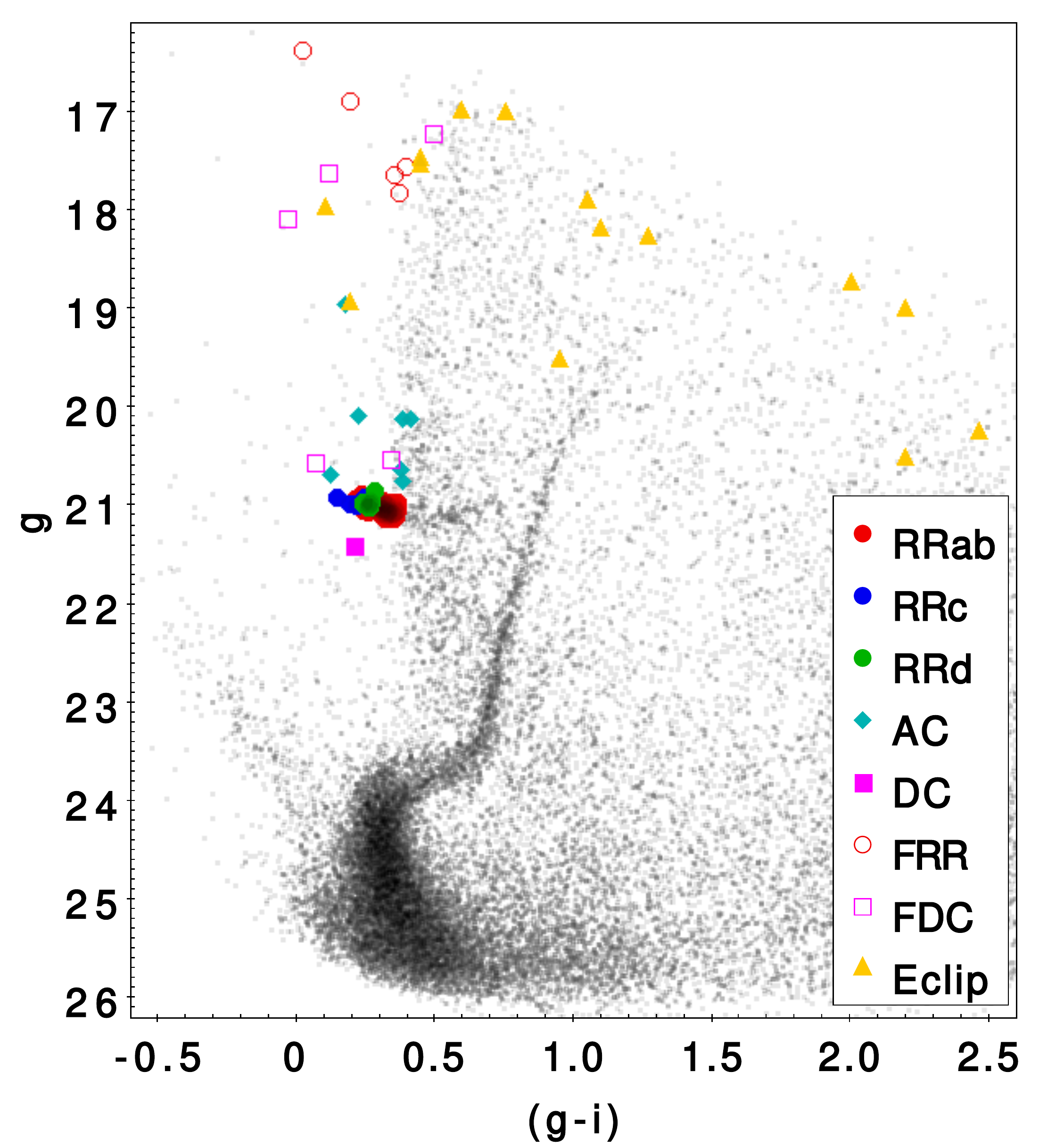}
\caption{Magnitude g versus color (g-i) of stars in Crater II (gray background) within $1\times r_h$ of the center of the galaxy from the deep photometry in the companion paper \citep{walker19} . The periodic variable stars (colored symbols) include stars found in the full FoV (130 stars). Legend is the same as in Figure~\ref{fig:Sky}.}
\label{fig:CMD}
\end{figure*}

Candidate variable stars, of which there are 1,584, were searched for periodicity in the range 0.03 to 1.5 days using a multi-band version of the \citet{lafler65} string length method, as explained in \citet{vivas16}. Potential periodic variables \citep[those stars with the $\Lambda>2.0$, see][]{lafler65} were visually inspected. For those stars, we examined the phased light curves of the three best periods given by this method in order to detect possible spurious periods due to aliasing or harmonics. Classification of the periodic variable stars was done based on the shape of the light curve, period, amplitude, and position on the CMD (Figure~\ref{fig:CMD}). To better see the features of Crater II in the CMD, we only plot in Figure~\ref{fig:CMD} stars within a circle of $31\arcmin$ from the center (that is, enclosing an area of $1r_h$). However, all variable stars within the full FoV are displayed in the figure. This CMD comes from the deep photometry made on stacked images as described in our companion paper \citep{walker19}.

We classified 130 periodic variables, of which 103 are RR Lyrae stars, 7 are anomalous Cepheids, 6 are dwarf Cepheids, and 14 are eclipsing binaries of different types. This means an increase of 36 periodic variables with respect to the findings by \citet{joo18}. Although the majority of the variable stars seem to be associated with Crater II based on their location in the CMD, there are some stars that most likely belong to the field foreground population. Those field variables are mostly located in the outskirts of the FoV (Figure~\ref{fig:Sky}). Coordinates, periods, amplitudes and mean magnitudes in $g$ and $i$, color excess E(B-V)$_{\rm SFD}$ from \citet{schlegel98} (which are meant to be used with the coefficients by \citet{schlafly11} given in \S~\ref{sec:ext} to obtain the extinctions), and classification are provided in Table~\ref{tab:var}. The mean magnitudes of RR Lyrae stars of type ab and c come from integrating the best fitted light curve template as explained in \S~\ref{sec:rrl}. For the rest of the variable stars, the mean magnitudes are phase-weighted intensity-averaged magnitudes calculated following the recipe provided by \citet{saha90}.

\begin{figure*}
\centering
\includegraphics[width=0.4\textwidth]{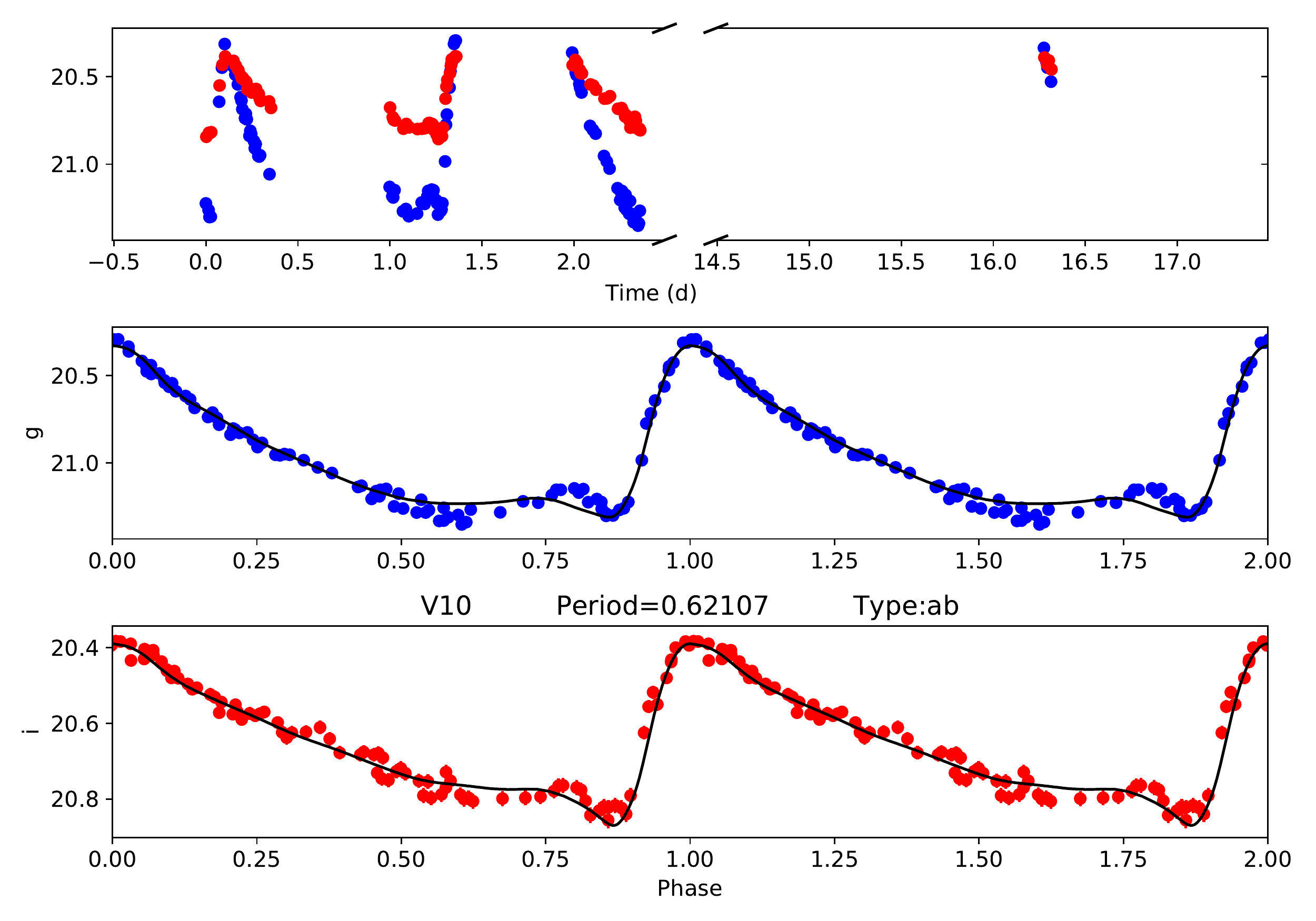}
\includegraphics[width=0.4\textwidth]{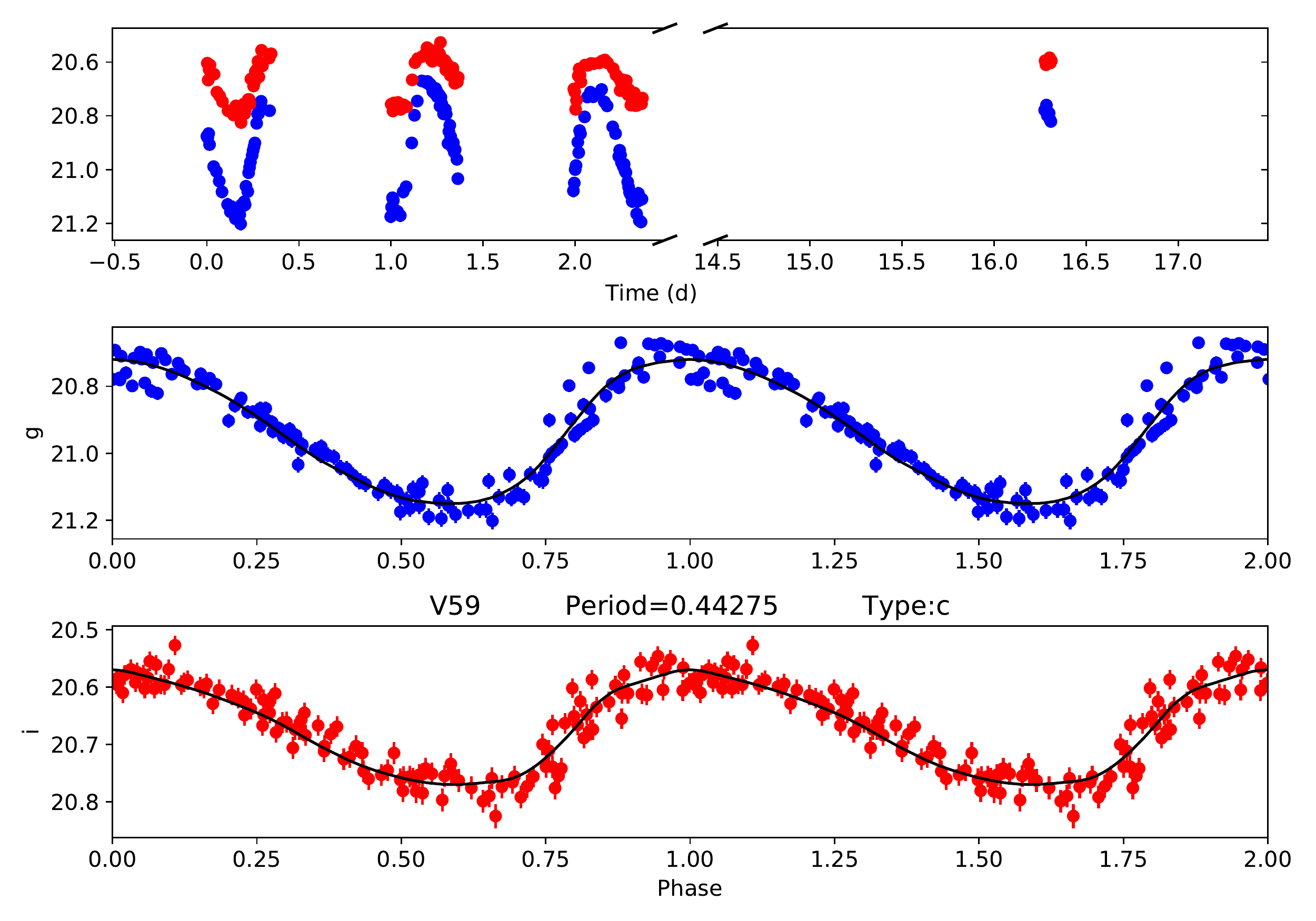}
\includegraphics[width=0.4\textwidth]{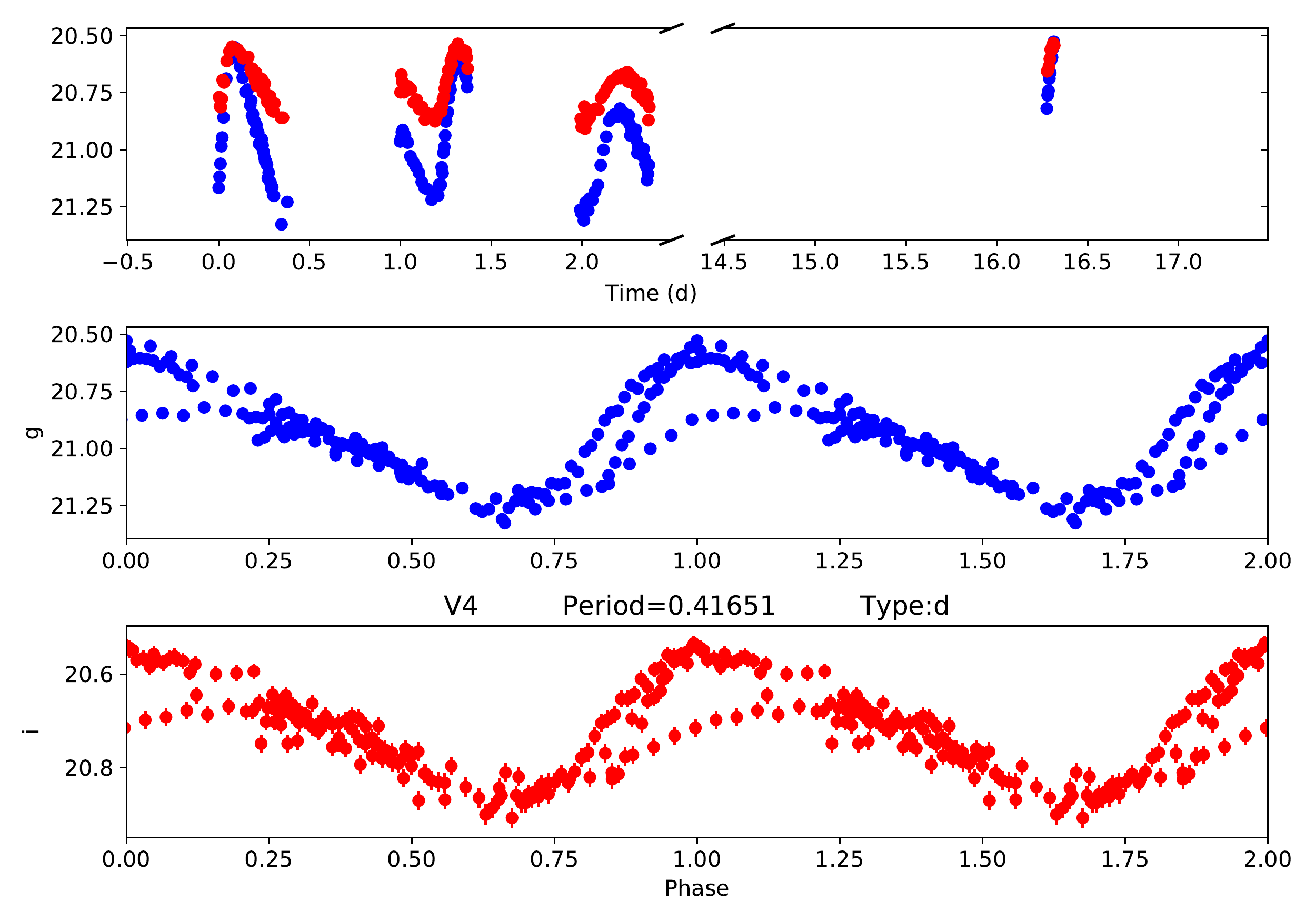}
\includegraphics[width=0.4\textwidth]{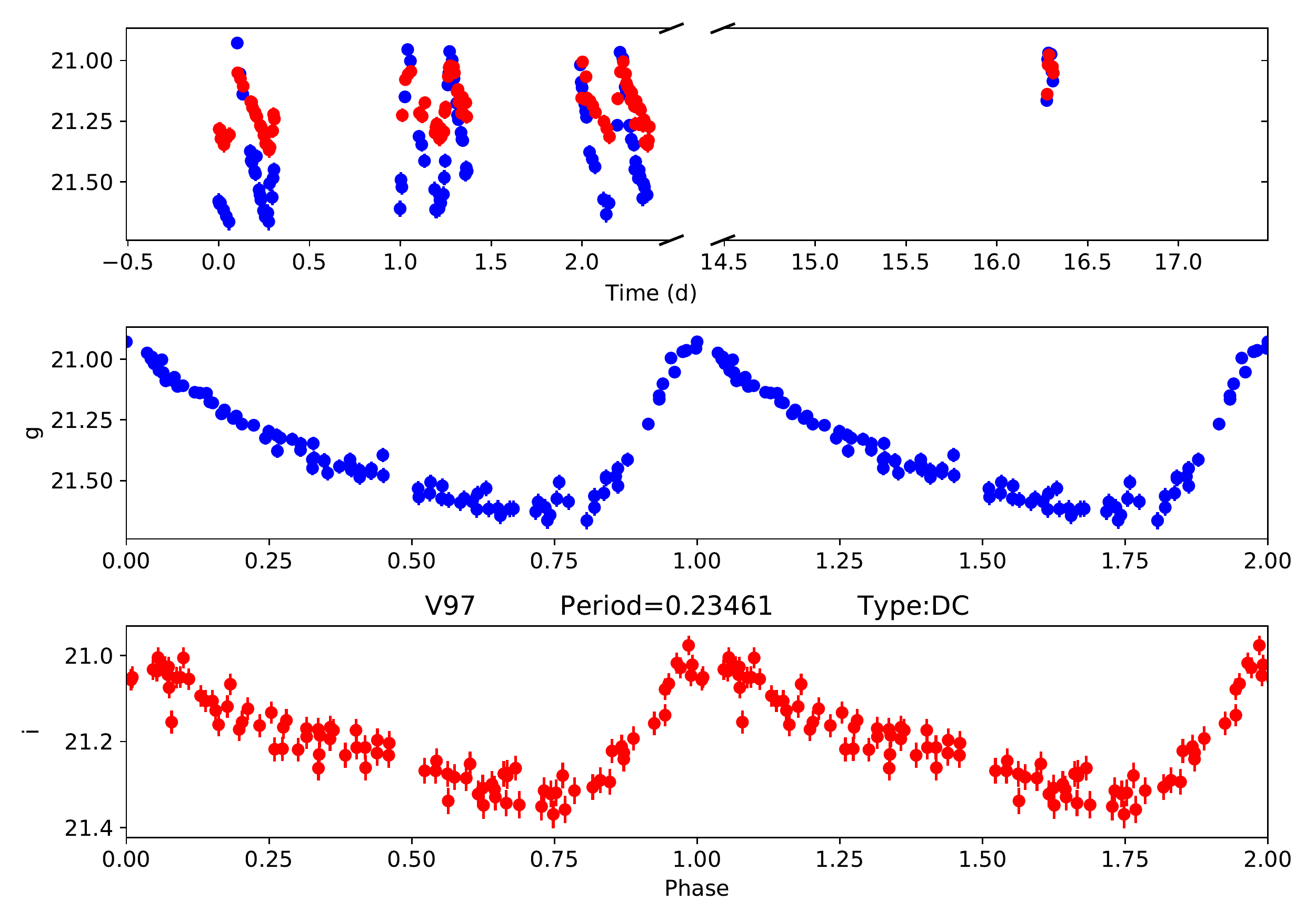}
\caption{Examples of light curves of RR Lyrae stars of type ab (V10), type c (V59), and type d (V4). We also show the lightcurve of the dwarf Cepheid star V97. Magnitudes in $g$ and $i$ are shown in blue and red, respectively. The top panel for each star is the time series in days from the time of the first observation. The two lower panels show the phased light curve. The phased lightcurves of RR Lyrae stars of types ab and c also display the best fitted template from the library by \citet{sesar10} as a black solid line. The complete set of lightcurves is available in the online version of the Journal.}
\label{fig:lc}
\end{figure*}

We recovered 94 out of 97 stars in \citet{joo18}. The three stars that were not recovered are V24, V79 and V96, which are all outside our footprint. The last 2 are classified by \citet{joo18} as field RR Lyrae stars of the type ab. V24 would be the only Crater II RR Lyrae star not in our data. We list all variables in Table~\ref{tab:var} using the ID convention followed by \citet{joo18} and \citet{monelli18}. Our periods and classification may differ from those in previous works for some stars. Notes on individual variables and differences in classification and/or period are discussed in the Appendix (\S~\ref{sec:appendix}). New periodic variable stars are listed in Table~\ref{tab:var}, starting with V99. Light curves are shown in Figure ~\ref{fig:lc} and companion online supplementary material, and the individual epoch photometry for all periodic variable stars is available from Table~\ref{tab:lc}.

The variable stars found in Crater II are in general too faint for Gaia DR2 \citep{gaia18}. Except for the faintest star in our sample (the DC star), all have a Gaia DR2 counterpart, although proper motions are not available for all stars and errors are large ($>1.5$ mas/yr) for the bulk of our variables. Thirteen of our stars are marked with the {\sl phot\_variable\_flag}, which indicates that they are indeed variable stars. These 13 stars include all of the field RR Lyrae stars, four anomalous Cepheids, and four of the Crater II RR Lyrae stars. Only the five field RR Lyrae stars are among the 140,784 RR Lyrae stars in table {\sl gaiadr2.vari\_rrlyrae} \citep{clementini19}, with agreement in their periods with ours. Our classification for one of those stars, V111, is however different (see \S~\ref{sec:appendix}).

\begin{table*}
\scriptsize
\caption{Periodic Variable Stars in Crater II}
\label{tab:var}
\begin{tabular}{cccccccccccc}
\hline
ID  & $\alpha$(J2000.0) & $\delta$ (J2000.0) & Period & $N_g$ & $\Delta_g$ & $\langle g \rangle$ &  $N_i$ & $\Delta_i$ & $\langle i \rangle$ & E(B-V)$_{\rm SFD}$ & Type$^{a}$ \\
      & (deg)                    & (deg)                      & (d)       &           & (mag)          & (mag)                      &            & (mag)        & (mag)                      &  (mag) &        \\
\hline
  V1 &    177.19560 &    -18.62472 &   0.76343 &  157 &   1.12 &   20.76 &  158 &   0.61 &   20.37 & 0.034 & AC     \\
  V2 &    177.24786 &    -18.16946 &   0.60442 &  125 &   1.17 &   21.08 &  125 &   0.51 &   20.78 & 0.032 & ab     \\
  V3 &    177.16149 &    -18.52788 &   0.60078 &  156 &   1.06 &   21.00 &  157 &   0.53 &   20.73 & 0.033 & ab     \\
  V4 &    177.21320 &    -18.27621 &   0.41651 &  155 &   0.80 &   21.00 &  158 &   0.37 &   20.73 & 0.032 & d      \\
  V5 &    177.21616 &    -18.53250 &   0.59900 &  152 &   1.29 &   21.08 &  157 &   0.61 &   20.75 & 0.033 & ab     \\
  V6 &    177.26842 &    -18.03395 &   0.40436 &  157 &   0.70 &   21.03 &  158 &   0.33 &   20.77 & 0.034 & d      \\
  V7 &    177.12839 &    -17.69956 &   0.59112 &  125 &   0.68 &   20.89 &  124 &   0.33 &   20.66 & 0.029 & ab     \\
  V8 &    177.19582 &    -18.52521 &   0.64982 &  157 &   0.63 &   21.00 &  158 &   0.26 &   20.72 & 0.033 & ab     \\
  V9 &    176.50267 &    -18.82727 &   0.59209 &   92 &   1.35 &   21.05 &   94 &   0.64 &   20.77 & 0.039 & ab     \\
 V10 &    176.96547 &    -18.18674 &   0.62107 &   90 &   0.98 &   20.93 &   92 &   0.48 &   20.64 & 0.036 & ab     \\
 V11 &    177.19876 &    -17.96955 &   0.67417 &  154 &   0.69 &   21.00 &  158 &   0.34 &   20.66 & 0.031 & ab     \\
 V12 &    177.17085 &    -18.52933 &   0.64053 &  157 &   0.52 &   21.06 &  158 &   0.27 &   20.73 & 0.033 & ab     \\
 V13 &    177.21498 &    -18.69282 &   0.64841 &  156 &   0.75 &   20.97 &  158 &   0.34 &   20.66 & 0.035 & ab     \\
 V14 &    177.16512 &    -18.15677 &   0.60836 &   91 &   0.69 &   21.05 &   87 &   0.35 &   20.74 & 0.032 & ab     \\
 V15 &    176.98760 &    -18.14437 &   0.63772 &   65 &   0.87 &   21.01 &   63 &   0.41 &   20.68 & 0.036 & ab     \\
 V16 &    177.14011 &    -18.07523 &   0.60167 &  158 &   0.82 &   21.01 &  157 &   0.34 &   20.73 & 0.032 & ab     \\
 V17 &    176.89240 &    -18.65649 &   0.62346 &  123 &   0.83 &   21.07 &  123 &   0.39 &   20.74 & 0.041 & ab     \\
 V18 &    176.88820 &    -18.10849 &   0.62832 &  155 &   0.73 &   21.05 &  157 &   0.36 &   20.72 & 0.038 & ab     \\
 V19 &    177.01731 &    -18.07904 &   0.63046 &  158 &   0.82 &   21.04 &  158 &   0.40 &   20.73 & 0.035 & ab     \\
 V20 &    177.14748 &    -18.51524 &   0.62306 &   89 &   0.79 &   21.02 &   92 &   0.37 &   20.72 & 0.034 & ab     \\
 V21 &    177.27399 &    -18.53421 &   0.63792 &  157 &   0.58 &   21.12 &  157 &   0.28 &   20.79 & 0.033 & ab     \\
 V22 &    177.06211 &    -18.25257 &   0.59772 &  155 &   1.27 &   20.99 &  156 &   0.61 &   20.73 & 0.033 & ab     \\
 V23 &    177.04428 &    -18.29525 &   0.61414 &  155 &   1.01 &   21.04 &  154 &   0.47 &   20.72 & 0.033 & ab     \\
 V25 &    177.27432 &    -18.67974 &   0.60938 &   91 &   0.66 &   21.08 &   92 &   0.35 &   20.78 & 0.035 & ab     \\
 V26 &    177.93807 &    -18.49433 &   0.77390 &  123 &   0.76 &   20.65 &  125 &   0.35 &   20.27 & 0.034 & AC     \\
 V27 &    177.60977 &    -17.98442 &   0.64729 &   97 &   0.77 &   20.98 &   99 &   0.37 &   20.66 & 0.034 & ab     \\
 V28 &    177.56619 &    -18.16064 &   0.62493 &  122 &   0.88 &   21.04 &  125 &   0.41 &   20.75 & 0.033 & ab     \\
 V29 &    177.34787 &    -18.33577 &   0.41992 &  124 &   0.75 &   20.97 &  124 &   0.37 &   20.70 & 0.032 & d      \\
 V30 &    178.06358 &    -17.99346 &   0.42438 &  113 &   0.68 &   20.99 &  117 &   0.44 &   20.75 & 0.032 & d      \\
 V31 &    177.93793 &    -18.12586 &   0.62490 &  157 &   0.64 &   20.97 &  157 &   0.31 &   20.65 & 0.032 & ab     \\
 V32 &    177.62062 &    -18.27896 &   0.60514 &  155 &   0.63 &   21.05 &  158 &   0.31 &   20.74 & 0.032 & ab     \\
 V33 &    177.44019 &    -18.54901 &   0.42092 &  158 &   0.79 &   21.03 &  158 &   0.38 &   20.77 & 0.033 & d      \\
 V34 &    177.68722 &    -18.58766 &   0.42521 &  157 &   0.64 &   20.98 &  158 &   0.33 &   20.70 & 0.037 & d      \\
 V35 &    177.48457 &    -18.62868 &   0.63159 &  158 &   0.82 &   20.96 &  157 &   0.40 &   20.66 & 0.034 & ab     \\
 V36 &    177.75522 &    -18.53922 &   0.68097 &  156 &   0.43 &   21.03 &  157 &   0.24 &   20.68 & 0.037 & ab     \\
 V37 &    178.26864 &    -17.92536 &   0.59212 &  150 &   0.61 &   20.95 &  151 &   0.24 &   20.70 & 0.034 & ab     \\
 V38 &    177.67416 &    -18.67226 &   0.62239 &   91 &   0.88 &   21.10 &   91 &   0.42 &   20.78 & 0.036 & ab     \\
 V39 &    177.37063 &    -18.02628 &   0.64217 &  125 &   0.71 &   21.00 &  133 &   0.36 &   20.67 & 0.032 & ab     \\
 V40 &    177.44287 &    -18.69602 &   0.62839 &  155 &   0.77 &   21.10 &  158 &   0.43 &   20.76 & 0.034 & ab     \\
 V41 &    177.51493 &    -18.74165 &   0.61848 &  154 &   0.63 &   21.08 &  155 &   0.29 &   20.75 & 0.035 & ab     \\
 V42 &    177.49996 &    -18.09225 &   0.62583 &  154 &   0.40 &   21.14 &  158 &   0.20 &   20.81 & 0.034 & ab     \\
 V43 &    177.67095 &    -18.27519 &   0.62583 &  156 &   0.81 &   20.97 &  158 &   0.40 &   20.68 & 0.033 & ab     \\
 V44 &    177.46115 &    -18.75020 &   0.64879 &  158 &   0.72 &   21.06 &  158 &   0.34 &   20.72 & 0.035 & ab     \\
 V45 &    177.65084 &    -17.95820 &   0.61624 &   95 &   0.96 &   20.99 &   93 &   0.47 &   20.70 & 0.033 & ab     \\
 V46 &    177.40338 &    -18.59078 &   0.61605 &  158 &   1.06 &   21.00 &  158 &   0.55 &   20.73 & 0.034 & ab     \\
 V47 &    177.61706 &    -17.87743 &   0.62079 &  144 &   0.49 &   21.11 &  145 &   0.25 &   20.79 & 0.033 & ab     \\
 V48 &    177.62306 &    -18.71747 &   0.61971 &   93 &   0.36 &   21.09 &   95 &   0.20 &   20.76 & 0.035 & ab     \\
 V49 &    177.53316 &    -18.69624 &   0.61542 &  158 &   0.71 &   21.06 &  158 &   0.33 &   20.74 & 0.034 & ab     \\
 V50 &    177.44466 &    -18.53698 &   0.58993 &  158 &   1.32 &   20.94 &  158 &   0.67 &   20.73 & 0.033 & ab     \\
 V51 &    177.55489 &    -18.99419 &   0.60827 &  124 &   0.80 &   21.04 &  124 &   0.41 &   20.74 & 0.043 & ab     \\
 V52 &    177.41907 &    -18.31799 &   0.63052 &   99 &   1.03 &   21.01 &   98 &   0.52 &   20.75 & 0.032 & ab     \\
 V53 &    177.60459 &    -18.73507 &   0.57260 &  154 &   1.18 &   20.99 &  156 &   0.63 &   20.78 & 0.036 & ab     \\
 V54 &    178.07123 &    -18.58137 &   0.58272 &  157 &   1.11 &   21.00 &  156 &   0.62 &   20.75 & 0.035 & ab     \\
 V55 &    177.95038 &    -18.62934 &   0.55380 &   94 &   1.00 &   21.01 &   96 &   0.48 &   20.79 & 0.036 & ab     \\
 V56 &    177.43688 &    -18.58405 &   0.56658 &  158 &   1.36 &   20.97 &  157 &   0.69 &   20.74 & 0.034 & ab     \\
 V57 &    177.47367 &    -18.27769 &   0.63231 &  124 &   0.79 &   21.14 &  124 &   0.38 &   20.82 & 0.031 & ab     \\
 V58 &    177.52640 &    -18.76454 &   0.57348 &  157 &   1.03 &   21.06 &  158 &   0.52 &   20.82 & 0.035 & ab     \\
 V59 &    177.31118 &    -18.76809 &   0.44275 &  124 &   0.43 &   20.92 &  124 &   0.20 &   20.67 & 0.036 & c      \\
 V60 &    176.79041 &    -18.37111 &   0.62627 &  157 &   0.78 &   21.05 &  158 &   0.35 &   20.74 & 0.037 & ab     \\
 V61 &    177.29405 &    -18.01777 &   0.61888 &   61 &   0.83 &   21.00 &   63 &   0.43 &   20.71 & 0.033 & ab     \\
 V62 &    177.59517 &    -18.41398 &   0.61351 &  154 &   0.69 &   20.98 &  158 &   0.34 &   20.71 & 0.034 & ab     \\
 V63 &    177.62824 &    -18.35910 &   0.63779 &  155 &   0.80 &   20.93 &  158 &   0.41 &   20.64 & 0.033 & ab     \\
 V64 &    177.57774 &    -18.41142 &   0.42994 &  152 &   0.78 &   21.02 &  156 &   0.40 &   20.76 & 0.034 & d      \\
 V65 &    176.93770 &    -18.39088 &   0.65029 &  156 &   0.83 &   21.00 &  158 &   0.42 &   20.69 & 0.035 & ab     \\
 V66 &    177.44579 &    -18.45585 &   0.62121 &  158 &   0.91 &   21.03 &  158 &   0.46 &   20.75 & 0.032 & ab     \\
 V67 &    177.28468 &    -18.80382 &   0.57829 &   99 &   1.27 &   20.99 &   98 &   0.71 &   20.79 & 0.037 & ab     \\
 V68 &    177.30212 &    -18.32688 &   0.65272 &  125 &   0.53 &   21.09 &  123 &   0.27 &   20.74 & 0.031 & ab     \\
 V69 &    177.67654 &    -18.76187 &   0.72365 &  158 &   0.37 &   21.06 &  158 &   0.18 &   20.69 & 0.038 & ab     \\
 V70 &    177.30013 &    -18.26403 &   0.62234 &  156 &   0.89 &   21.01 &  156 &   0.42 &   20.71 & 0.034 & ab     \\
 V71 &    177.56625 &    -18.43157 &   0.60581 &  155 &   1.00 &   21.01 &  158 &   0.50 &   20.75 & 0.034 & ab     \\
 V72 &    177.20828 &    -18.47438 &   0.65598 &   97 &   0.40 &   21.09 &   99 &   0.20 &   20.75 & 0.033 & ab     \\
 V73 &    177.37589 &    -18.69234 &   0.64523 &  158 &   0.56 &   21.07 &  156 &   0.29 &   20.74 & 0.034 & ab     \\
 V74 &    177.51637 &    -18.43635 &   0.63405 &  155 &   0.74 &   21.08 &  157 &   0.36 &   20.76 & 0.033 & ab     \\
 V75 &    177.52497 &    -18.40928 &   0.60753 &  152 &   0.93 &   21.03 &  157 &   0.47 &   20.75 & 0.033 & ab     \\
 V76 &    177.26700 &    -18.15219 &   0.65131 &  100 &   0.44 &   21.13 &   99 &   0.24 &   20.77 & 0.032 & ab     \\
 V77 &    177.32238 &    -18.64888 &   0.61299 &   66 &   0.98 &   21.00 &   65 &   0.48 &   20.72 & 0.034 & ab     \\
 V78 &    177.32042 &    -18.56514 &   0.60142 &   95 &   1.01 &   21.07 &   95 &   0.51 &   20.82 & 0.033 & ab     \\
 V80 &    177.96476 &    -18.51461 &   0.44678 &   58 &   0.70 &   20.86 &   59 &   0.34 &   20.58 & 0.034 & d      \\
\hline
\hline
\end{tabular}
\end{table*} 
 
\begin{table*}
\scriptsize
\contcaption{Periodic Variable Stars in Crater II}
\label{tab:continued}
\begin{tabular}{cccccccccccc}
\hline
ID  & $\alpha$(J2000.0) & $\delta$ (J2000.0) & Period & $N_g$ & $\Delta_g$ & $\langle g \rangle$ &  $N_i$ & $\Delta_i$ & $\langle i \rangle$ & E(B-V) & Type$^{a}$ \\
      & (deg)                    & (deg)                      & (d)       &           & (mag)          & (mag)                      &            & (mag)        & (mag)                      &  (mag) &        \\
\hline 
 V81 &    176.78436 &    -18.87495 &   0.72716 &  157 &   0.52 &   20.98 &  158 &   0.25 &   20.62 & 0.042 & ab     \\
 V82 &    176.78820 &    -18.53957 &   0.41775 &  156 &   0.80 &   20.98 &  157 &   0.39 &   20.73 & 0.042 & d      \\
 V83 &    177.18941 &    -18.16634 &   0.62033 &  124 &   0.62 &   21.05 &  125 &   0.32 &   20.73 & 0.032 & ab     \\
 V84 &    176.67931 &    -17.81401 &   0.61621 &   99 &   0.97 &   21.02 &   99 &   0.48 &   20.71 & 0.035 & ab     \\
 V85 &    176.50947 &    -18.53670 &   0.61364 &  153 &   0.48 &   21.03 &  156 &   0.23 &   20.73 & 0.039 & ab     \\
 V86 &    177.73818 &    -18.76755 &   0.41224 &  157 &   0.77 &   20.69 &  158 &   0.39 &   20.57 & 0.039 & AC      \\
 V87 &    177.77935 &    -18.65725 &   0.65529 &  123 &   0.61 &   21.05 &  125 &   0.29 &   20.72 & 0.037 & ab     \\
 V88 &    177.44079 &    -17.64595 &   0.63010 &  151 &   0.76 &   20.98 &  144 &   0.36 &   20.66 & 0.032 & ab     \\
 V89 &    177.70780 &    -17.69896 &   0.62628 &  149 &   0.74 &   21.04 &  153 &   0.38 &   20.71 & 0.033 & ab     \\
 V90 &    178.02531 &    -18.73588 &   0.43233 &  158 &   0.73 &   20.99 &  157 &   0.35 &   20.72 & 0.037 & d      \\
 V91 &    177.86732 &    -18.74082 &   0.41601 &  158 &   0.56 &   21.01 &  157 &   0.27 &   20.80 & 0.040 & c      \\
 V92 &    177.24183 &    -18.36888 &   0.72508 &  152 &   0.41 &   20.97 &  158 &   0.20 &   20.62 & 0.031 & ab     \\
 V93 &    177.10197 &    -18.38634 &   0.65446 &  157 &   0.30 &   21.09 &  157 &   0.15 &   20.73 & 0.036 & ab     \\
 V94 &    177.32342 &    -18.10084 &   0.67200 &   95 &   0.25 &   21.07 &   95 &   0.12 &   20.71 & 0.033 & ab     \\
 V95 &    177.74326 &    -18.45078 &   0.61383 &  155 &   0.93 &   21.00 &  157 &   0.45 &   20.72 & 0.036 & ab     \\
 V97 &    177.31728 &    -18.30992 &   0.23461 &   97 &   0.74 &   21.42 &   94 &   0.39 &   21.21 & 0.032 & DC     \\
 V99 &    177.55409 &    -18.29873 &   0.66890 &  150 &   0.31 &   21.06 &  151 &   0.15 &   20.70 & 0.032 & ab     \\
V100 &    177.25352 &    -18.45239 &   0.66237 &  154 &   0.21 &   21.10 &  157 &   0.11 &   20.75 & 0.032 & ab     \\
V101 &    177.86770 &    -18.36617 &   0.61129 &  155 &   0.91 &   21.07 &  158 &   0.43 &   20.77 & 0.033 & ab     \\
V102 &    177.20003 &    -18.59608 &   0.66192 &  154 &   0.23 &   21.08 &  158 &   0.11 &   20.72 & 0.033 & ab     \\
V103 &    177.30564 &    -18.86134 &   0.65772 &  158 &   0.55 &   21.15 &  158 &   0.26 &   20.80 & 0.038 & ab     \\
V104 &    177.32655 &    -17.78051 &   0.37752 &  158 &   0.72 &   20.93 &  157 &   0.36 &   20.79 & 0.031 & c      \\
V105 &    177.41865 &    -17.80933 &   0.40889 &  152 &   0.50 &   20.99 &  154 &   0.23 &   20.77 & 0.032 & c      \\
V106 &    176.61045 &    -18.17140 &   0.40071 &  113 &   0.53 &   21.00 &  112 &   0.27 &   20.80 & 0.037 & c      \\
V107 &    177.41128 &    -18.07878 &   0.51347 &  156 &   0.45 &   20.12 &  158 &   0.18 &   19.74 & 0.034 & AC     \\
V108 &    176.75842 &    -18.30692 &   0.37040 &  100 &   0.58 &   18.97 &   99 &   0.26 &   18.79 & 0.038 & AC     \\
V109 &    176.95431 &    -18.28700 &   0.91334 &  156 &   0.87 &   20.14 &  157 &   0.37 &   19.72 & 0.034 & AC     \\
V110 &    177.26743 &    -18.72983 &   0.52200 &  157 &   0.77 &   20.10 &  158 &   0.37 &   19.87 & 0.036 & AC     \\
V111 &    177.79205 &    -17.44083 &   0.52217 &   89 &   1.54 &   17.57 &   89 &   0.76 &   17.17 & 0.036 & Fab    \\
V112 &    177.63233 &    -17.55664 &   0.47647 &   99 &   0.93 &   16.90 &  104 &   0.44 &   16.71 & 0.033 & Fab    \\
V113 &    178.10231 &    -18.45572 &   0.56774 &  155 &   1.06 &   17.66 &  158 &   0.51 &   17.31 & 0.035 & Fab    \\
V114 &    176.83038 &    -18.97353 &   0.30110 &   67 &   0.49 &   16.40 &   51 &   0.23 &   16.37 & 0.042 & Fc     \\
V115 &    177.83524 &    -19.01431 &   0.50537 &  148 &   1.43 &   17.83 &  153 &   0.71 &   17.46 & 0.040 & Fab    \\
V116 &    177.48901 &    -17.44548 &   0.18827 &   91 &   0.39 &   20.28 &   95 &   0.18 &   17.81 & 0.034 & Eclip  \\
V117 &    177.18708 &    -17.53108 &   0.23465 &   90 &   0.76 &   18.77 &   87 &   0.42 &   16.77 & 0.033 & Eclip  \\
V118 &    177.46704 &    -18.05752 &   0.27439 &  156 &   0.56 &   17.04 &   92 &   0.41 &   16.29 & 0.034 & Eclip  \\
V119 &    176.54502 &    -18.19017 &   0.13470 &  141 &   0.22 &   17.93 &  145 &   0.16 &   16.88 & 0.036 & Eclip  \\
V120 &    177.31113 &    -18.31283 &   0.26394 &  100 &   0.68 &   18.22 &   99 &   0.52 &   17.12 & 0.033 & Eclip  \\
V121 &    177.33753 &    -18.46950 &   0.15305 &  144 &   0.52 &   17.03 &  107 &   0.41 &   16.43 & 0.031 & Eclip  \\
V122 &    177.68939 &    -18.34611 &   0.85274 &   89 &   0.27 &   17.50 &   89 &   0.25 &   17.05 & 0.033 & Eclip  \\
V123 &    177.93489 &    -18.41856 &   0.77515 &  150 &   0.53 &   19.03 &  146 &   0.38 &   16.83 & 0.034 & Eclip  \\
V124 &    178.24466 &    -18.66277 &   0.12801 &  125 &   0.28 &   19.54 &  125 &   0.19 &   18.59 & 0.034 & Eclip  \\
V125 &    176.79710 &    -18.80953 &   0.58013 &   95 &   0.27 &   18.00 &   94 &   0.23 &   17.90 & 0.043 & Eclip  \\
V126 &    176.80808 &    -18.73907 &   0.15264 &  156 &   0.30 &   17.58 &  157 &   0.21 &   17.13 & 0.043 & Eclip  \\
V127 &    178.12052 &    -18.82907 &   0.26495 &   91 &   0.61 &   18.96 &   91 &   0.44 &   18.77 & 0.036 & Eclip  \\
V128 &    177.29142 &    -19.06658 &   0.39558 &  104 &   0.22 &   18.30 &  100 &   0.18 &   17.03 & 0.043 & Eclip  \\
V129 &    177.63314 &    -19.05527 &   0.30975 &   97 &   0.17 &   20.54 &  103 &   0.08 &   18.34 & 0.042 & Eclip  \\
V130 &    177.84632 &    -17.78471 &   0.05150 &  158 &   0.20 &   20.58 &  156 &   0.17 &   20.51 & 0.032 & FDC    \\
V131 &    176.41853 &    -18.01057 &   0.07399 &   91 &   0.39 &   17.63 &   92 &   0.17 &   17.51 & 0.031 & FDC    \\
V132 &    176.43775 &    -18.29070 &   0.12260 &  151 &   0.15 &   20.54 &  151 &   0.10 &   20.20 & 0.036 & FDC    \\
V133 &    176.52508 &    -18.21109 &   0.03820 &  158 &   0.18 &   18.10 &  158 &   0.08 &   18.13 & 0.035 & FDC    \\
V134 &    177.68287 &    -19.31769 &   0.14251 &   66 &   0.17 &   17.24 &   66 &   0.13 &   16.75 & 0.043 & FDC    \\
\hline
\hline
\end{tabular}

\begin{tablenotes}
\item $^a$Types: ab $=$ type ab RR Lyrae stars; c $=$ type c RR Lyrae stars; d $=$ type d RR Lyrae stars; AC $=$ anomalous Cepheids;  DC $=$ dwarf Cepheid stars; FDC $=$ Field dwarf Cepheid stars; Fab $=$ Field type ab RR Lyrae stars; Fc $=$ Field type c RR Lyrae stars; Eclip $=$ eclipsing binary stars.
\end{tablenotes}
\end{table*} 

\begin{table}
\small
\centering
\caption{Time Series Photometry of Variable Stars in the Field of View of Crater II}
\label{tab:lc}
\begin{tabular}{ccccc}
\hline
ID  & JD & Mag & Error & Filter \\
      & (d) & (mag) & (mag) &       \\
\hline
  V1  & 2457832.508222  & 20.986  & 0.022  & g \\
  V1  & 2457832.513068  & 20.994  & 0.022  & g \\
  V1  & 2457832.517926  & 21.005  & 0.022  & g \\
  V1  & 2457832.522774  & 20.988  & 0.022  & g \\
  V1  & 2457832.527616  & 20.989  & 0.022  & g \\
\hline
\hline
\end{tabular}

\begin{tablenotes}
\item Table~\ref{tab:lc} is published in its entirety as online supplementary material. A portion is shown here for guidance regarding its form and content.
\end{tablenotes}

\end{table} 

\section{RR Lyrae Stars} \label{sec:rrl}

\begin{figure*}
\centering
\includegraphics[width=1.0\columnwidth]{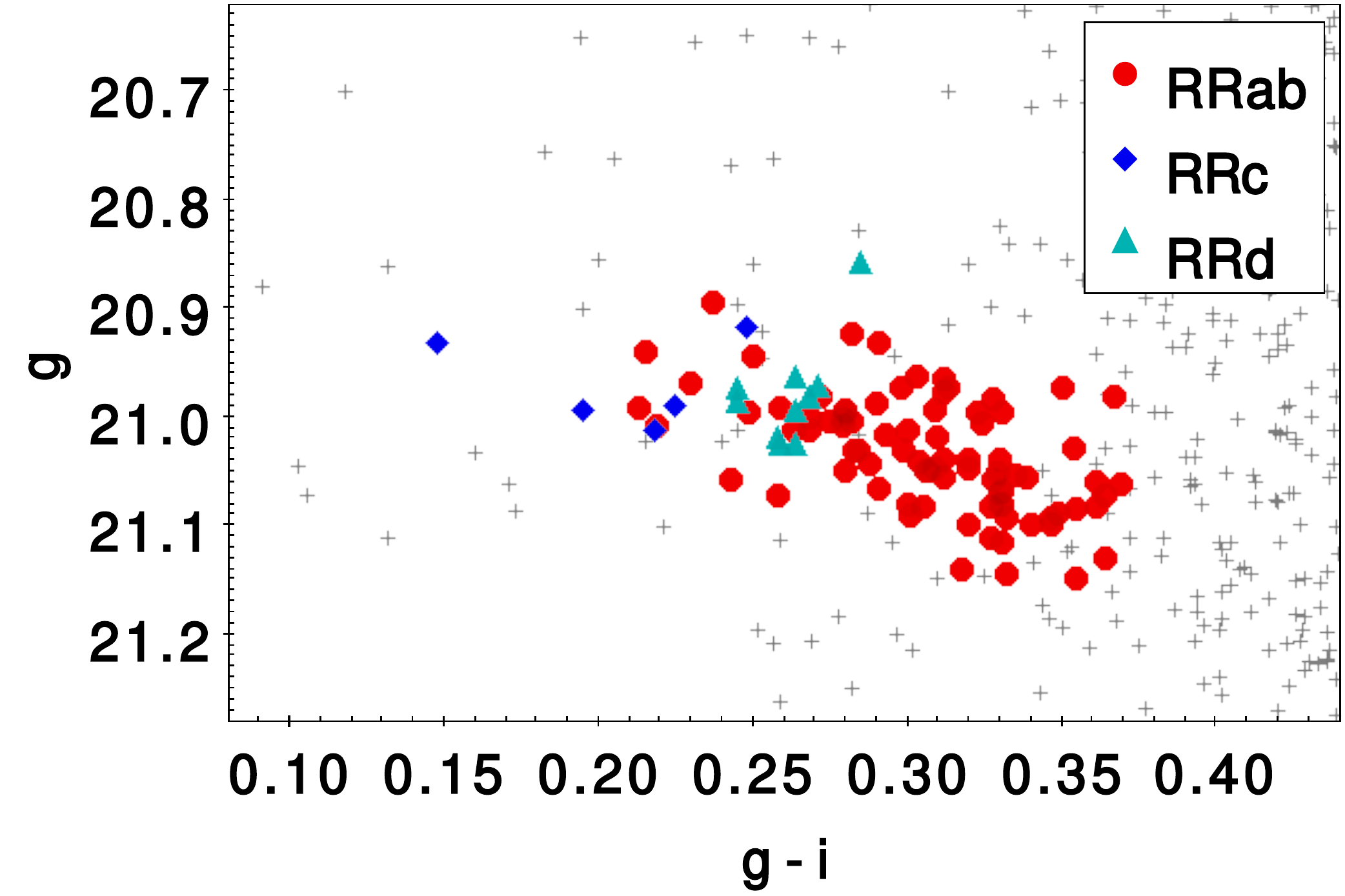}
\includegraphics[width=1.0\columnwidth]{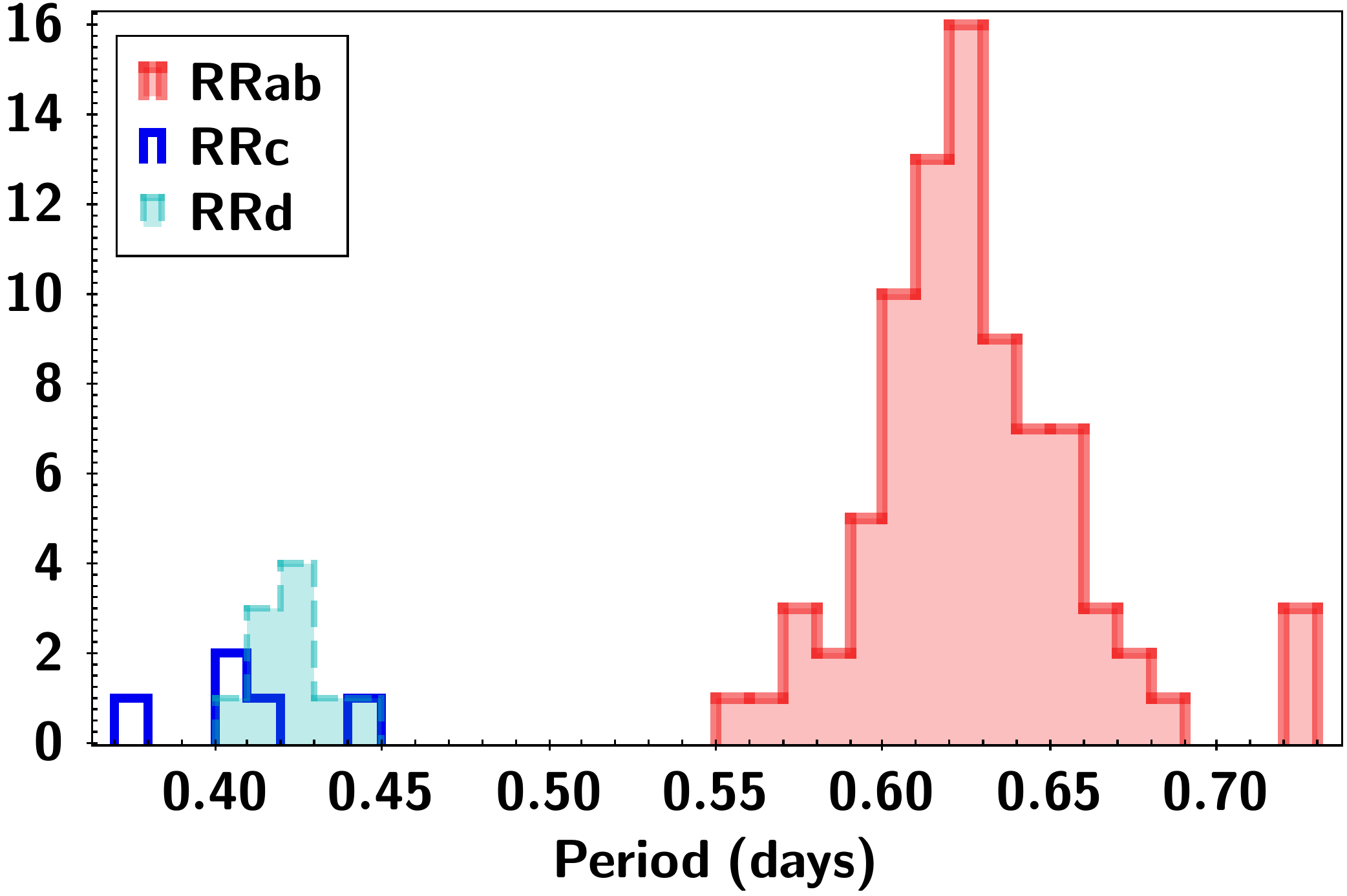}
\caption{(Left) A zoom of the CMD (Fig~\ref{fig:CMD}) in the region of the RR Lyrae stars. (Right) Period distribution of the RR Lyrae stars.}
\label{fig:CMDZoom}
\end{figure*}

Crater II is very rich in RR Lyrae stars. Out of the 103 RR Lyrae stars detected in this work, 98 are located on the horizontal branch of the galaxy (Figure~\ref{fig:CMD}), at $g\sim 21.0$. The remaining 5 stars (V111 to V115 in Table~\ref{tab:var}), are much brighter, with $g$ magnitudes between 16.4 and 17.7, and are likely field stars in the foreground of Crater II. As mentioned above, we missed one Crater II RR Lyrae star located beyond our footprint, which was measured by \citet{joo18}. Including this star, Crater II has a total of 99 RR Lyrae stars, of which 84 are type ab (\rrab), 5 are type c (\rrc) and 10 are type d (\rrd).
A zoom into the variables in the horizontal branch in the CMD can be seen in the left panel of Figure~\ref{fig:CMDZoom}. As expected, {\rrab} stars are located toward the red edge of the instability strip, while the hotter {\rrc } are on the bluest end of the horizontal branch. Type d stars are in between {\rrab} and \rrc. 

With a median of 155 epochs in each band, the RR Lyrae stars in this work are extremely well characterized. The possibility of aliasing periods is basically zero since we can check the continuous coverage of each star over several hours during each observed night (top panels in Figure~\ref{fig:lc}). At the magnitude of the RR Lyrae stars, the photometric errors of individual measurements in the light curves are (from Figure~\ref{fig:error}), $\sim 0.02$ mag in both bands.

In order to obtain better estimates of the amplitude and mean magnitude of the RR Lyrae stars, we fitted templates from the library by \citet{sesar10} to the lightcurves (see Figure~\ref{fig:lc}). The mean magnitudes reported in Table~\ref{tab:var} were obtained by integrating the templates in intensity units and transforming the result back to magnitudes. This procedure was done only for the types {\rrab } and \rrc. The high cadence and short total baseline of our data naturally excludes the Blazhko effect to be present in the lightcurves. Although Blazhko periods can be as short as 5 days, they are most commonly tens of days \citep{catelan15,benko14}. Indeed, none of our lightcurves has signs of Blazhko effect. 

It is well known that amplitudes of RR Lyrae stars are larger in bluer filters than in red ones. This is clearly seen in the lightcurves in Figure~\ref{fig:lc}. The mean observed amplitude ratio in our sample of 93 RR Lyrae stars (excluding the {\rrd } stars) is $\Delta g / \Delta i = 2.04$ ($\sigma = 0.13$). Thus, the amplitude in $g$ is about twice the one in the $i$ band. 

The period distribution of the RR Lyrae stars is shown in the right panel of Figure~\ref{fig:CMDZoom}. The separation between the {\rrab } and the {\rrcd } is quite evident in this histogram. The transition period between {\rrab } and \rrc, defined as the shortest period of the {\rrab } \citep{smith95}, is $P_{tr} = 0.55$ d. The lack of {\rrab } with periods shorter than 0.48 days indicates that this galaxy has no High Amplitude Short Period (HASP) variables (see also \S~\ref{sec:bailey}). HASP stars are commonly found in systems which are more metal rich than [Fe/H]$\sim -1.5$ \citep{fiorentino15}. Thus, the RR Lyrae stars' period distribution point towards a metal-poor old population in Crater II.

The mean periods of the RR Lyrae stars in Crater II are 0.627 d ($\sigma = 0.03$ d) and 0.410 d ($\sigma = 0.02$ d) for {\rrab } and \rrc, respectively. Including the {\rrd } stars in the later group, the mean period of the {\rrcd } stars is 0.409 d ($\sigma = 0.02$ d). The mean periods obtained here are basically identical to the ones obtained by \citet{joo18} (we discuss some differences in periods and classification with our sample for a few individual stars in \S~\ref{sec:appendix}), 0.621 d and 0.423 d for their {\sl bona-fide} sample. \citet{monelli18} found a slightly shorter mean period for the type ab stars, 0.617 d. Because they covered only the central part of the galaxy, we explore later in this paper if that small difference in the mean period may be due to radial gradients in the population of RR Lyrae stars (\S~\ref{sec:gradient}).

\subsection{Double-mode Pulsators}

A surprisingly large number (10 stars) of double mode pulsators was found among the sample of RR Lyrae stars, comprising 10\% of the total sample. Their IDs are V4, V6, V29, V30, V33, V34, V64, V80, V82 and V90. As is usually the case for these stars, the dominant pulsation mode is the first-overtone one, which was easily selected with our string-length algorithm for measuring the period. The light curves phased with that period show, however, amplitude variations (see for example star V4 in Figure~\ref{fig:lc}). Such variations cannot be regarded as the Blazhko effect since they occur on a night-to-night basis (see top panel for V4 in Figure~\ref{fig:lc}), which is a timescale too short for this effect. The observed behavior, on the other hand, resembles the lightcurves of those type d RR Lyrae stars that have continuous coverage by the Kepler space telescope \citep{molnar15}. Double mode pulsators are best identified using Fourier analysis. We re-analyzed the time series for these stars using a full least-squares periodogram program written by Dr. L.A. Balona, pre-whitened by the first-overtone frequency in order to search for the fundamental frequency. The main period found with this procedure agrees with our initial periods with only small variations (usually $< 0.01$ d). Secondary peaks, after whitening, were present for all stars in the sample, confirming these are indeed double mode pulsators. The precision of those periods is, however, compromised by the relatively short time baseline of our observations. More observations spanning several pulsation cycles will be needed to identify correctly the secondary periods. We report here (Table~\ref{tab:var}) only the main (first overtone) periods. 

\subsection{Bailey Diagram \label{sec:bailey}}

\begin{figure}
\centering
\includegraphics[width=1.0\columnwidth]{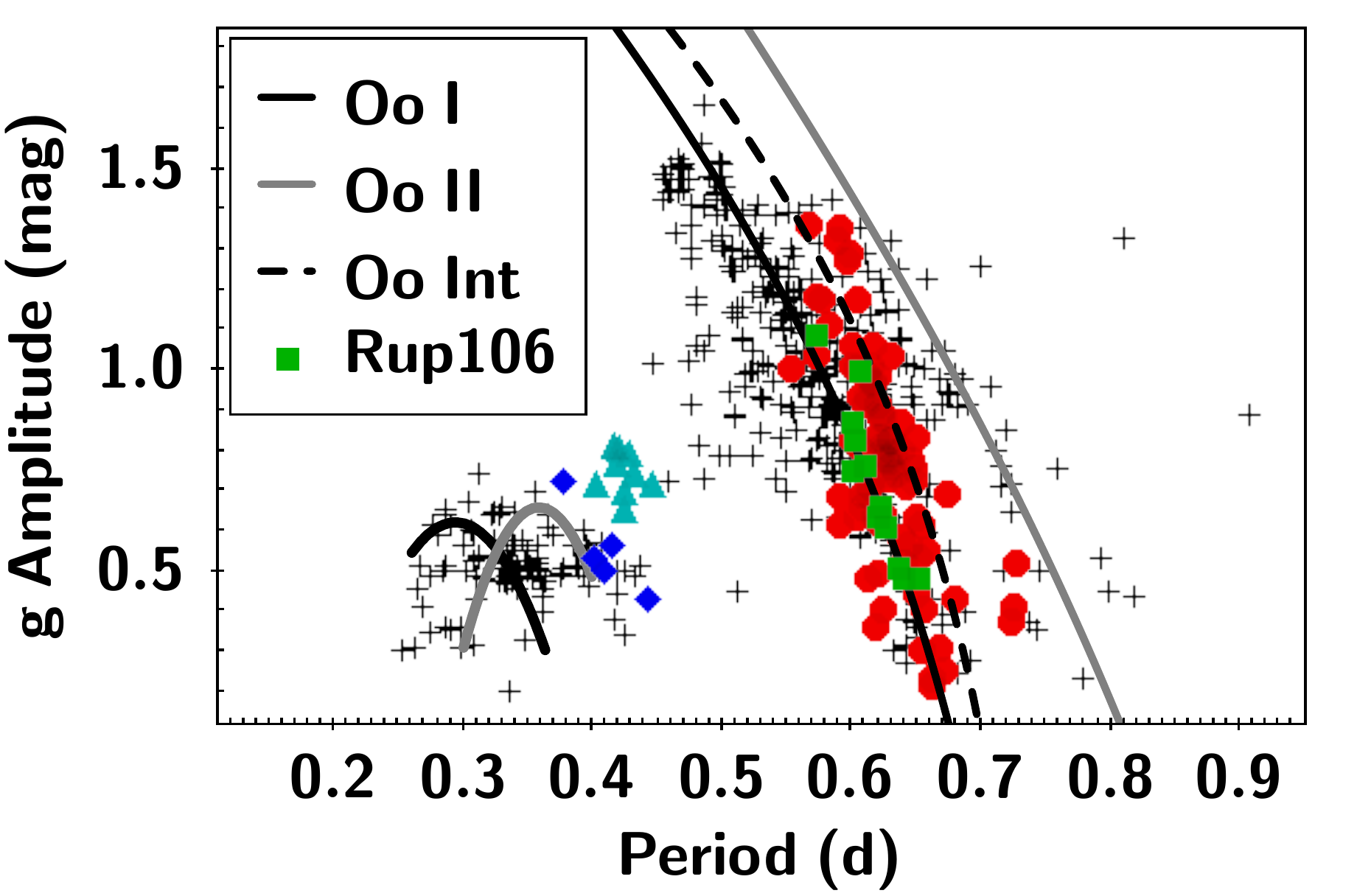}
\includegraphics[width=1.0\columnwidth]{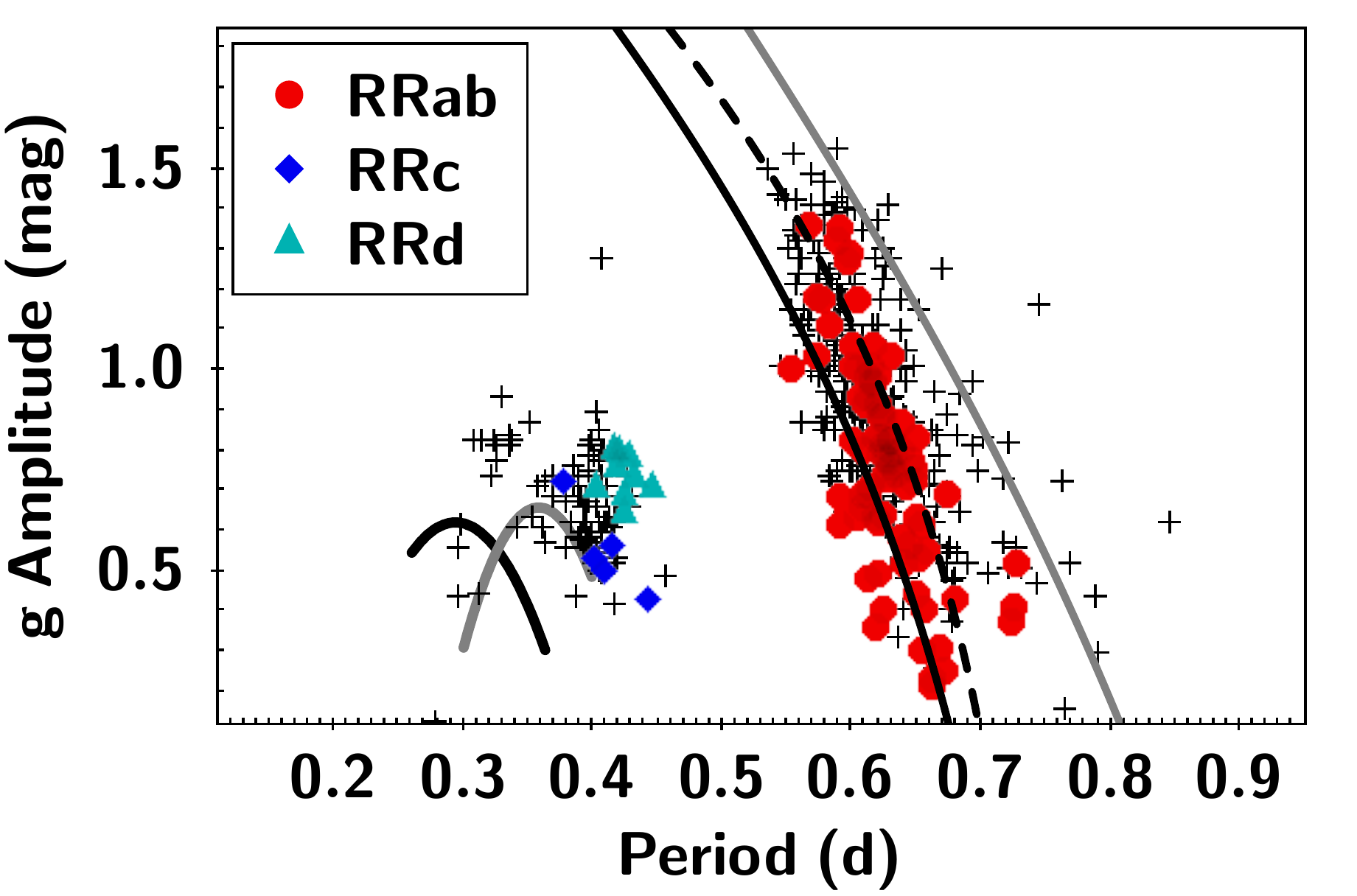}

\caption{(Top) Bailey Diagram (amplitude in $g$ versus period) of RR Lyrae stars in Crater II. For comparison, we show in the background (black $+$'s) the distribution of RR Lyrae stars in the Halo from SDSS Stripe 82 \citep{sesar10}. The ridge line of the Oo I, Oo II and Oo-int populations are shown as black and grey lines. The RR Lyrae stars in the globular cluster Rup 106 are shown as green squares.
(Bottom) Same as above but the background in this case are RR Lyrae stars in selected dwarf galaxies as compiled by \citet{braga16} and scaled to the $g$ band (see text).}
\label{fig:PA}
\end{figure}

The mean periods found in this work would indicate a classification as an Oosterhoff (Oo) II group \citep{oosterhoff39} since the nominal Oo II group has mean periods of 0.64d and 0.37d for {\rrab } and \rrc, respectively \citep[periods of Oo I groups are shorter, 0.55d and 0.32d,][]{smith95}. The transition period ($P_{tr}=0.55$ d) also agrees with an Oo II classification \citep{smith95}. However, Oo II groups are expected to have high numbers of {\rrc } stars ($N_c/N_{ab+c} = 0.44$), which is not seen here. Including the double-mode pulsators, our data indicate that Crater II has a ratio of only $15/(83+15)=0.15$. This ratio is similar to a Oo I population (expected to be 0.17). The small number of {\rrc } in this galaxy can be due, however, to the fact that the horizontal branch stops within the instability strip, and hence the galaxy does not produce many of the hotter type c variables. No blue horizontal branch (BHB) stars are present in this galaxy as seen in Figure~\ref{fig:CMD}. This morphology is consistent with a relatively young age for the HB stars, 11-12 Gyrs, as discussed in more detail in the companion paper \citep{walker19}.

Further insight on the Oo classification can be obtained from a Bailey Diagram (amplitude versus period), which is shown in Figure~\ref{fig:PA}. The Crater II RR Lyrae stars clearly separate in this diagram in {\rrab } (Period $\gtrsim 0.5$ d), and {\rrcd} (Period $\lesssim 0.5$ d). The {\rrab } have a rather vertical distribution in this diagram. For reference, we show the ridge line of the Oo populations. The ridgelines for the ab-types are taken from \citet{fabrizio19}, while for the c-types they were obtained from \citet{arellano15} and \citet{kunder13} for Oo I and Oo II, respectively. In all cases, the ridgelines were scaled from the $V$ amplitude in which they were defined to $g$ amplitude by multiplying by 1.29. This value of $A_V/A_g$ is based on the average of 472 stars in common between the SDSS Stripe 82 RR Lyrae stars \citep{sesar10}, which were measured in several filters including $g$, and those from the Catalina Real-Time Transient Survey \citep[CRTS,][]{drake14}, which have $V$ photometry. The lower amplitude ($\Delta g<0.7$ mag) {\rrab } stars in Crater II follow the Oo I line, but the stars shift toward longer periods for the higher amplitude variables. This behavior is different from what is seen in the field (top panel of Figure~\ref{fig:PA}), which contains abundant stars with large amplitudes and short periods, following the Oo I line. The field stars in this figure are RR Lyrae stars from SDSS Stripe 82 which encompass halo stars between $\sim 5-100$ kpc \citep{sesar10}. On the other hand, the {\rrcd } in Crater II fill only the long period tail for this class. As mentioned above, one of the reasons for this is that the horizontal branch ends within the instability strip, and hence does not produce many type c stars.

The distribution of Crater II RR Lyrae stars in the Bailey Diagram, on the other hand, is more similar to the overall distribution of stars in some of the Milky Way dwarf spheroidal galaxies. From the compilation of RR Lyrae stars in dwarf galaxies made by \citet{braga16}, we selected Canes Venatici I, Ursa Major I, Bootes I, Draco and Carina. Again, the $V$ amplitudes in that compilation were scaled to $g$ amplitudes. The distribution of the RR Lyrae stars in those galaxies in the Bailey diagram is pretty similar to the ones in Crater II. All of them are also missing the short period, high amplitude {\rrab } variables seen in the halo population. The {\rrc } are also clumped toward the Oo II locus, although with high dispersion. From the Bailey diagram it is clear that the Crater II \rrab, as well as the ones in dwarf galaxies, are mostly located in the Oo-intermediate region. The ridgeline of that population (dashed line), as determined by \citet{fabrizio19}, gets closer to the Oo I region at lower amplitudes, which is the behavior observed in Crater II. 

\citet{monelli18} previously found that the only globular cluster with a horizontal branch morphology similar to Crater II is Ruprecht (Rup) 106. This cluster has a red horizontal branch and several RR Lyrae stars, all of them being {\rrab } \citep[][July 2016 update]{kaluzny95,clement01}. As in Crater II, the lack of {\rrc } is due to the truncation of the horizontal branch within the instability strip. Rup 106 has a relatively young age \citep{dotter11}, which also seems to be the case for Crater II \citep{walker19}. The mean period of its 12 RR Lyrae stars is 0.616 d, very similar to the one for Crater II. Although this mean period usually corresponds to Oo II groups, the distribution in the Bailey diagram is actually closer to the Oo I ridgeline (Figure~\ref{fig:PA}, top), and has been given this classification \citep{clement01}. Indeed, the Rup 106 RR Lyrae stars occupy a similar locus as the Crater II stars in the Bailey diagram although its stars are closer to the OoI locus.

All the above suggest that Crater II is neither Oo I nor Oo II. 

\subsection{Period-Luminosity Relationship}\label{sec:pli}

\begin{figure}
\centering
\includegraphics[width=1.0\columnwidth]{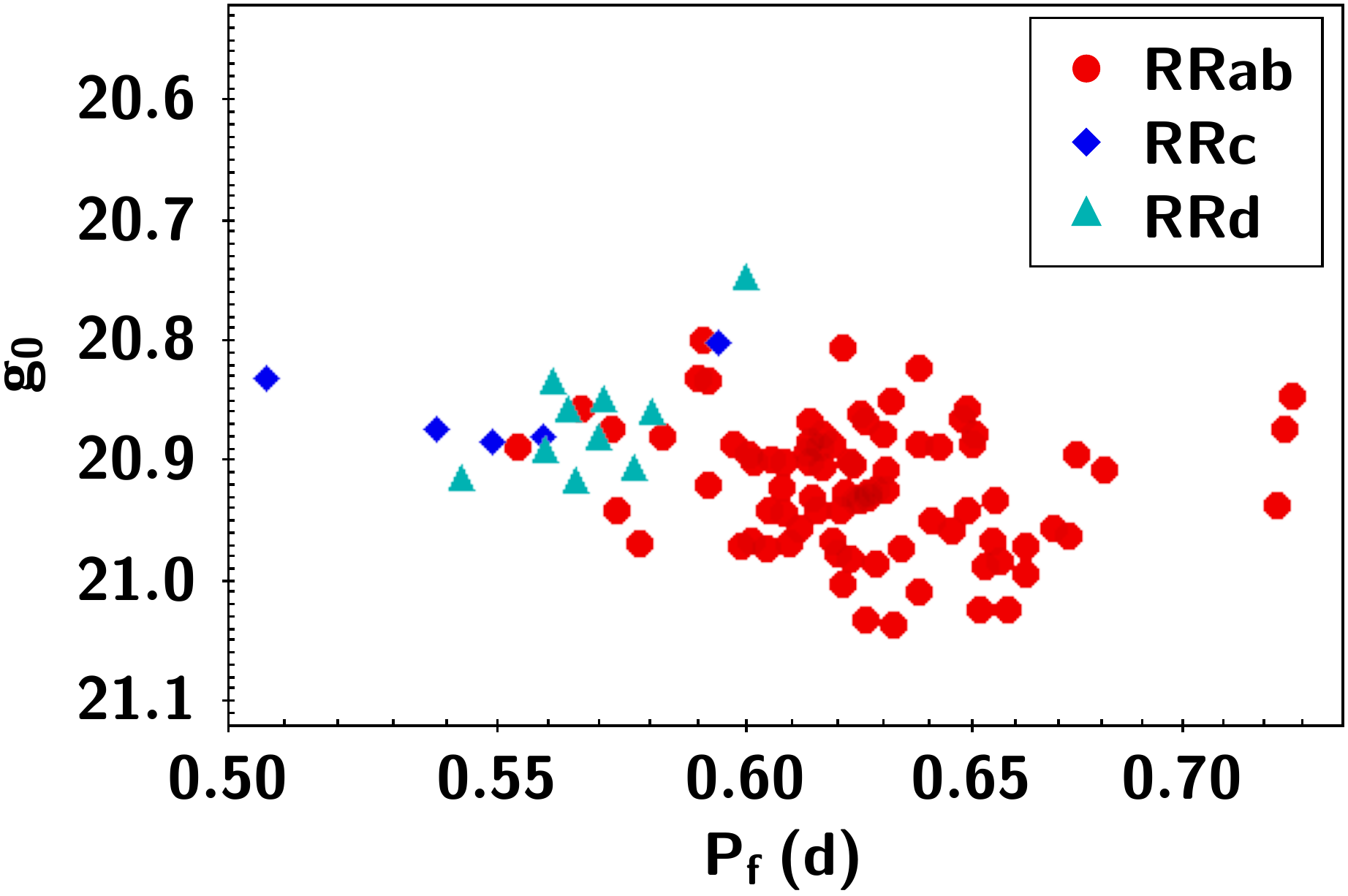}
\includegraphics[width=1.0\columnwidth]{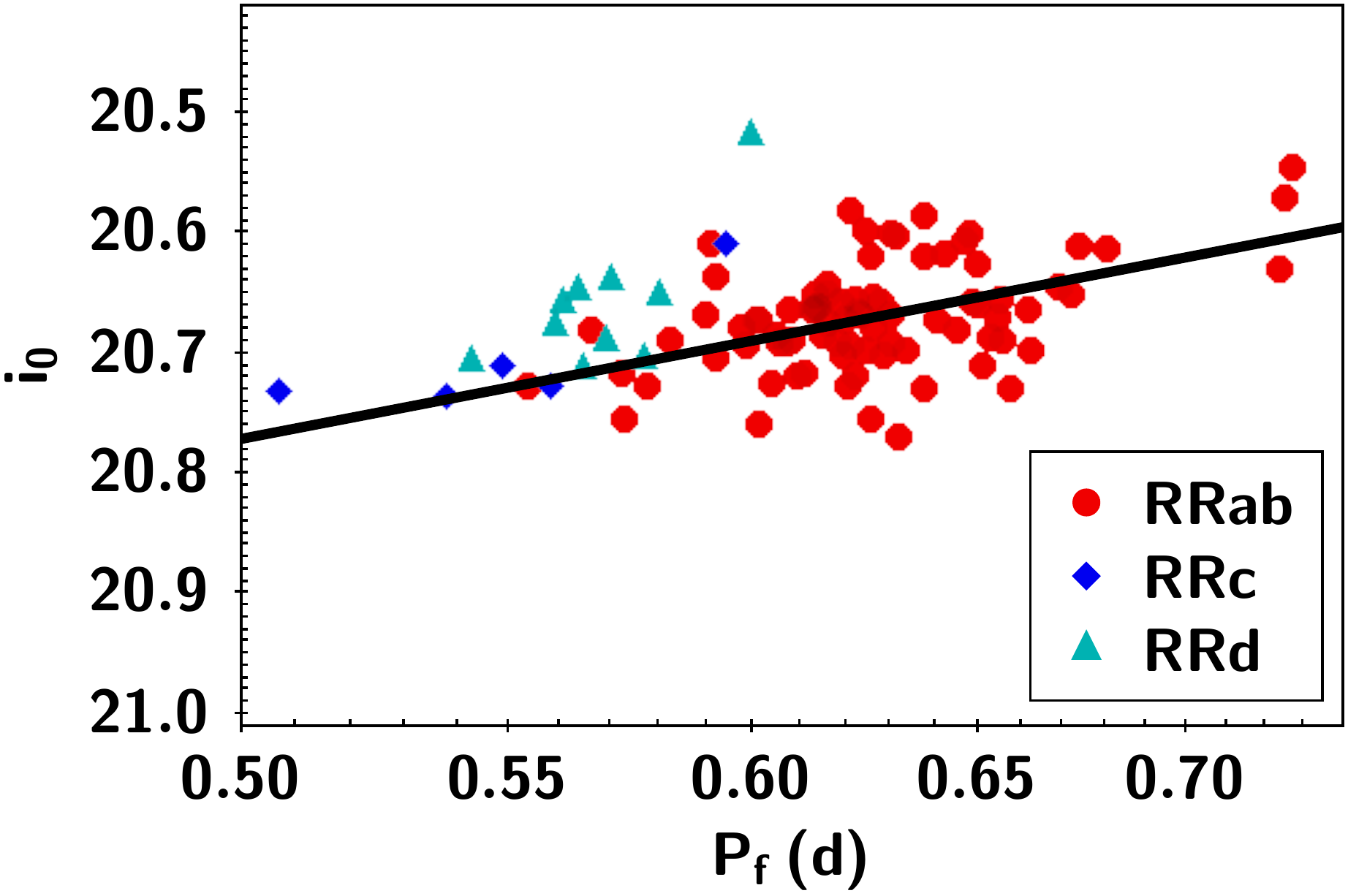}
\caption{Extinction-corrected magnitudes, $g_0$ (top) and $i_0$ (bottom) of RR Lyrae stars in Crater II as a function of the fundamental period $P_f$, on a logarithmic scale. The solid line in the bottom panel is not a fit to our data. It shows the PL relationship in $i$ from \citet{caceres08}, shifted to a distance modulus $\mu_0=20.333$ mag. Notice that the vertical span in magnitudes in both panels is the same.}
\label{fig:PL}
\end{figure}

As shown in Figure~\ref{fig:CMD} and ~\ref{fig:CMDZoom} (left), the RR Lyrae stars in Crater II are tightly clumped in the horizontal branch. The mean extinction-corrected magnitudes for all (98) RR Lyrae stars are $20.91 \pm 0.06$ and $20.67 \pm 0.05$ in $g_0$ and $i_0$, respectively. Although these dispersions are relatively small, we can minimize them by exploring dependences with period and with type.

In Figure~\ref{fig:PL} we show the extinction-corrected magnitudes, $g_0$ and $i_0$, as a function of the fundamental period (in logarithmic scale). The periods of types $c$ and $d$ were adjusted to the fundamental period using $\log P_f = \log P + 0.128$ \citep{catelan09}. These figures show some well known behavior: {\sl (i)} RR Lyrae stars exhibit a greater dispersion in their mean magnitudes in the $g$ band than in the $i$ band \citep{caceres08}; {\sl (ii)} there is a tight correlation of $i_0$ with period (a Period-Luminosity, PL, relationship), while no such dependence on period exists in the $g$ band; and {\sl (iii)} {\rrcd } exhibit slightly brighter magnitudes in both bands when compared with {\rrab } \citep[see also][]{marconi05,vivas17}.

Because of the above, we estimate the distance to Crater II using the $i$-band mean magnitudes of only \rrab, which are in any case very numerous in this galaxy. We calculate distances to all the {\rrab } using the Period-Luminosity-Metallicity (PLZ) relationship in \citet{caceres08}:

\begin{equation}
M_i = 0.908 - 1.035 \log{P} + 0.220 \log{Z}
\label{eq:PLi}
\end{equation}

\noindent
where

\begin{equation}
\log{Z} = {\rm [Fe/H]} + \log{(0.638 \times 10^{[\alpha/\rm{Fe}]} + 0.362)}- 1.765
\end{equation}

The formal uncertainty in equation~\ref{eq:PLi} is 0.045 mag. We assume a metallicity of [Fe/H]$=-2.0$, which is in agreement with spectroscopic measurements by both \citet{caldwell17} and \citet{fu19}. For the $\alpha$ enhancement, $[\alpha/\rm{Fe}]$, we used $+0.3$ dex, which is a value appropriate for the old population of dwarf galaxies  \citep{pritzl05}.

The mean distance modulus of {\rrab } in Crater II is $\mu_0 = 20.333 \pm 0.004$, where the quoted error is the error of the mean of the distribution. The standard deviation of the distance moduli is 0.039 mag, while the systematic error is 0.068 mag. The latter was estimated by error propagation including the uncertainty in eq~\ref{eq:PLi}, 0.045 mag, a metallicity dispersion of 0.17 dex (see~\ref{sec:metallicity}), an uncertainty of 0.1 dex in [$\alpha$/Fe], a 10\% error in interstellar extinction \citep{schlegel98}, and a photometric error of 0.02 mag, which is the mean error for stars at the level of the horizontal branch of Crater II (see Figure~\ref{fig:error}). In Figure~\ref{fig:PL} (bottom) we show equation~\ref{eq:PLi} shifted by a distance modulus of 20.333 to show that the PLZ relationship used here is adequate and its slope correctly describes our data. The resulting distance modulus is in excellent agreement with the one derived by \citet{monelli18}, $\mu_0 = 20.30 \pm 0.08$ using a smaller sample of RR Lyrae stars, a different method (the Period-Wesenheit relation), and a different photometric system to derive distances. Our distance is slightly larger, but still consistent within errors, than the one obtained by \citet{joo18}, $\mu_0=20.25 \pm 0.10$, probably due in part to the difference in metallicity assumed by those authors for the RR Lyrae stars ([Fe/H]=$-1.65$). Our distance modulus translates into a distance of $116.5^{+3.7}_{-3.6} \pm 0.2$ kpc to Crater II, where the errors are the systematic and statistical ones, respectively.

The small dispersion in the distance modulus of the {\rrab } of Crater II, 0.039 mag, leaves little room for a dispersion in metallicities among the RR Lyrae star population in this galaxy. Hints of this can be obtained from the dispersion observed in globular clusters, which are almost all believed to have no dispersion in [Fe/H]. \citet{vivas17} measure PL relationships in the DECam passbands for the globular cluster M5. In the $i$-band, the standard deviation of the PL relationship was 0.02 mag. This leaves 0.033 mags when subtracted in quadrature as possible due to a dispersion in metallicity in the RR Lyrae stars. We discuss this with more detail in next section.

\subsection{Metallicity Distribution \label{sec:metallicity}}

\begin{figure}
    \centering
    \includegraphics[width=0.95\columnwidth]{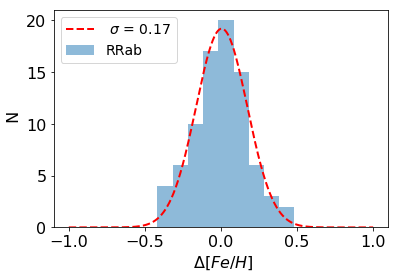}
    \caption{$\Delta$[Fe/H] distribution of the RRab stars in Crater~II.}
    \label{fig:mdf}
\end{figure}

In a first approach, the metallicity dispersion can be obtained by just dividing the dispersion in the distance modulus by 0.22, the coefficient of $\log Z$ in eq~\ref{eq:PLi}, since the contribution of the errors in the periods are negligible. This results in a metallicity dispersion of 0.18 dex.

More formally, it is possible to obtain independent measurements of the metallicity of the RR Lyrae population when there are photometric data available in several bands, for which well defined Period-Luminosity-Metallicity (PLZ) relationships exist \citep{martinez2016a,bono19}. In our case, we only have 2 bands and only $i$ has a well defined PLZ relationship \citep{caceres08}. Thus, we cannot get an independent measurement of [Fe/H]. However, by inverting the PLZ (equation~\ref{eq:PLi}), we can obtain the  metallicity distribution of RR Lyrae stars, possibly affected by a global zero point shift, that is dependent on the mean [Fe/H] assumed to derive the distance modulus. The [Fe/H] dispersion around the mean value can tell us about the spread in metallicity of the population of RR Lyrae stars in Crater II.

Figure~\ref{fig:mdf} shows the $\Delta$[Fe/H] distribution obtained for the {\rrab} stars of Crater~II, defined as [Fe/H]$_i$-$\langle \rm{[Fe/H]} \rangle$, where $i$ represents each RR Lyrae star. The $\Delta$[Fe/H] distribution is single-peaked and is well represented by a Gaussian of $\sigma$=0.17 dex. There are no signs of multiple stellar populations with different metallicities from this distribution. This result is in agreement with \citet{caldwell17} who measured a metallicity dispersion of 0.22 dex from 62 red giant stars members of Crater II, and with \citet{fu19} who obtained a 0.18 dex metallicity dispersion based on independent spectroscopic observations of red giant stars. 

Using this method of inverting the PLZ, \citet{martinez2016a} estimated the expected metallicity dispersion from a mono-metallic population by studying the globular cluster Reticulum, which is believed to have no multiple stellar populations. They obtained $\Delta$[Fe/H]$=0.25$ dex when using all RR Lyrae stars in the cluster and $\Delta$[Fe/H]$=0.07$ dex when using only the lightcurves with the best quality. Since the Crater II lightcurves are all high quality, the second value may be the most relevant here. Thus, subtracting this value in quadrature from the dispersion derived from Figure~\ref{fig:mdf}, we obtain a maximum real metallicity spread of 0.17 dex. This means that the spread in the metallicity of the old population of Crater~II is very small, but not completely negligible. Our comparison with the dispersion of the PL in the globular cluster M5 (\S~\ref{sec:pli}) is also in agreement with a small, but non-negligible spread in metallicity in Crater II.

\subsection{Spatial Distribution \label{sec:sp}}

\begin{table*}
\small
\caption{Spatial Distribution Parameters}
\label{tab:sp}
\begin{tabular}{lccccc}
\hline
Sample & RA & DEC & $\epsilon$ & PA & 1-$\sigma$ Semi-major axis \\
       & (deg) & (deg) &       & (deg) & (arcmin)                \\
\hline
\citet{torrealba16} & 177.310 & -18.413 & $<0.1$ & ... & ... \\
RR Lyrae stars      & 177.37 $\pm$ 0.03  & -18.42 $\pm$ 0.01 & 0.24 $\pm$ 0.04 & 153 $\pm$ 7 & 21.3 $\pm$ 0.8 \\
GB photometric members & 177.328 $\pm$ 0.003 & -18.418 $\pm$ 0.005 & 0.12 $\pm$ 0.02 & 135 $\pm$ 4 & 23.1 $\pm$ 0.3 \\ 
\hline
\hline
\end{tabular}
\end{table*} 

\begin{figure}
\includegraphics[width=1.0\columnwidth]{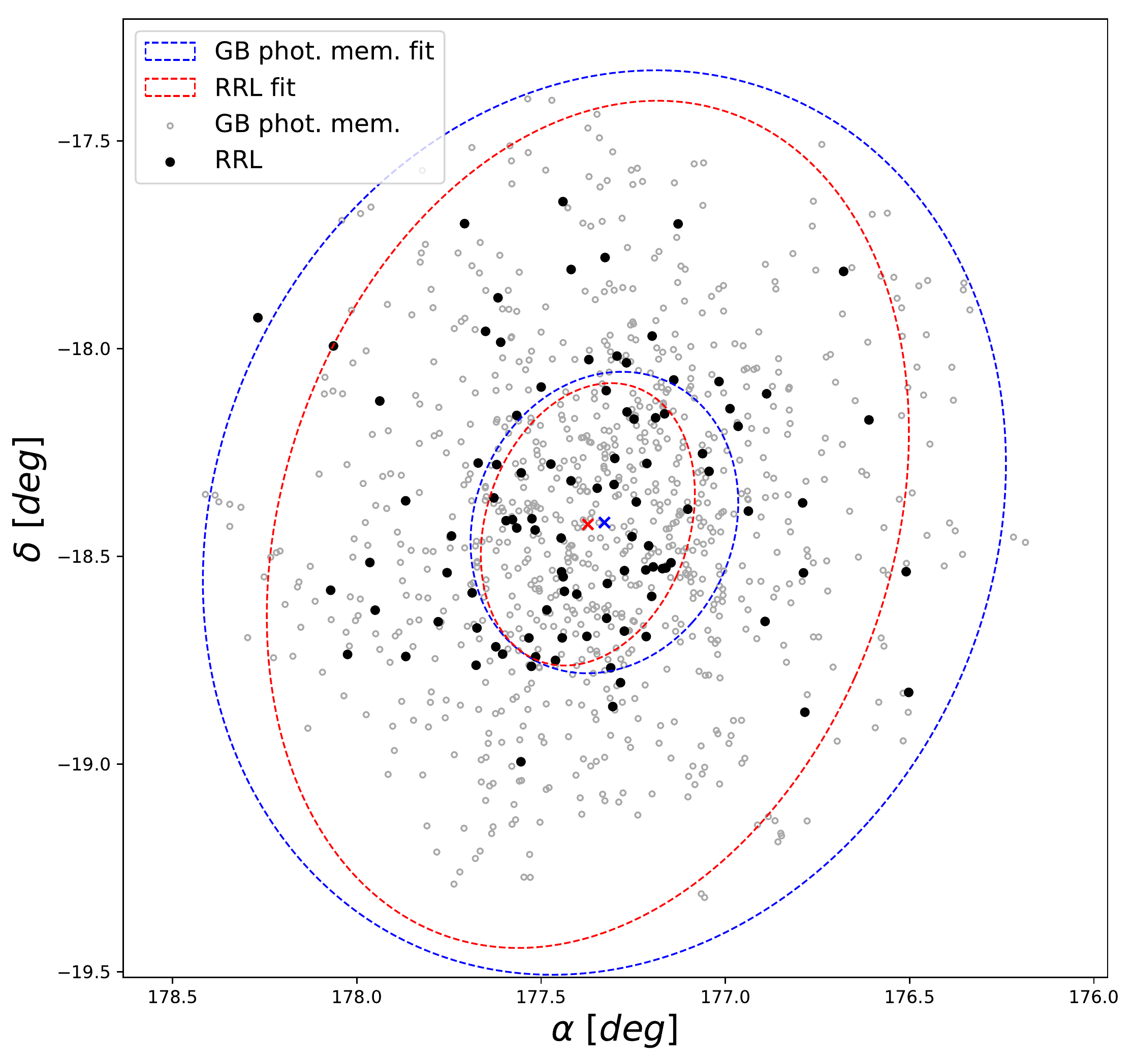}
\caption{Spatial distribution of RR Lyrae stars (solid circles) and high probability giant branch photometric members of Crater II (open circles). Ellipses are the best fitted $1\sigma$ and $3\sigma$ bi-variate Gaussian distribution to each population. The $\times$ symbols indicate the centers of each set of ellipses.}
\label{fig:sp}
\end{figure}

The number of RR Lyrae stars in Crater II is high enough that the stars can provide information on the spatial distribution of the old population in the galaxy. To do this, we computed the components of a bivariate Gaussian distribution to the coordinates ($\alpha,\delta$) of the RR Lyrae variable stars, using the routine {\texttt fit\_bivariate\_normal} in AstroML \citep{ivezic14}. The result is seen in Figure~\ref{fig:sp}, and the corresponding parameters together with their uncertainties are provided in Table~\ref{tab:sp}. To estimate the uncertainties we followed a bootstrapping method. We randomly selected 90\% of the RR Lyrae stars and derived the ellipse parameters of this reduced sample. This procedure was repeated 10,000 times. The errors reported in Table~\ref{tab:sp} are the standard deviation of the fitted parameters in those 10,000 experiments.

The RR Lyrae stars in Crater II have a rather elongated distribution with an ellipticity of $e=0.24$. This result is in contrast with estimates provided in the discovery paper \citep{torrealba16} which suggested a more circular distribution of stars. The advantage of calculating the ellipticity of the galaxy using RR Lyrae stars is that no contamination by foreground stars is present. Halo RR Lyrae stars would be very rare at this large distance \citep{medina18}, and the chances of detecting one or more in such a small area of the sky are minimal. The different inferred shapes therefore, may be attributed to the purity of our sample.

The elongated shape of the RR Lyrae star population is supported by the distribution of 863 giant branch (GB) photometric members of the galaxy, selected as described in the companion paper \citep[][their section 3.2]{walker19}. This sample is more numerous than the RR Lyrae stars, but it is more prone to suffer contamination by foreground Milky Way stars because this is a galaxy with very low surface brightness. Using the same methodology described above, we also found an elongated distribution of stars, with similar center, size and position angle as the RR Lyrae stars, as shown by the blue ellipses in Figure~\ref{fig:sp}. The ellipticity, however, is not as large as for the RR Lyrae stars. Small differences between the parameters found with both samples are present (Table~\ref{tab:sp}). Those differences may arise from some contamination in the photometric sample, although we expect it to be quite low, of the order of $5\%$ \citep{walker19}. The RR Lyrae and GB stars may also preferentially trace different stellar populations which may have slightly different spatial distributions.

\subsection{Radial Profile and Gradients}\label{sec:gradient}

\begin{figure}
    \centering
    \includegraphics[width=0.95\columnwidth]{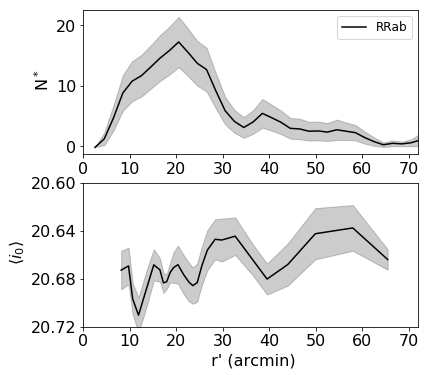}
    \caption{\textit{Top}. Running average of the number of {\rrab } stars as a function of the elliptical distance ($r'$). Shaded region show the Poisson uncertainties. \textit{Bottom}. Running average of the extinction-corrected $i$ magnitude, $\langle i_0 \rangle$, of the {\rrab} stars as a function of the elliptical distance. Shaded region show the standard deviation in each box.}
    \label{fig:Nab_mag_prof}
\end{figure}

\begin{figure}
    \centering
    \includegraphics[width=0.95\columnwidth]{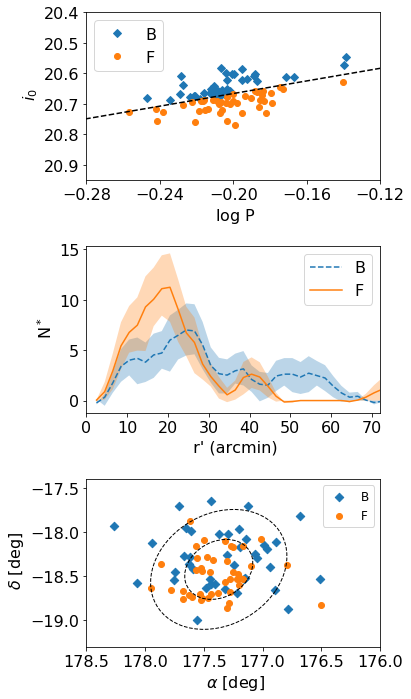}
     \caption{\textit{Top}. Dashed line represents the PL relation in the $i_0$-band (equation~\ref{eq:PLi}) at the distance of Crater II. The line is used to separate the bright (B) and Faint (F) samples of {\rrab } represented with blue diamonds and orange circles, respectively. \textit{Middle}. Running average of the number of bright and faint {\rrab } as a function of the elliptical distance ($r'$). Shaded regions show the Poisson uncertainties of these distributions.\textit{Bottom}. Spatial distribution of the two groups of {\rrab } stars: bright (blue diamonds) and faint (orange dots). Ellipses are the $1\sigma$ and 2$\sigma$ bivariate fits to the RR Lyrae star distribution (Table~\ref{tab:sp}).}
    \label{fig:BF_profile}
\end{figure}

Taking advantage of the sizable sample of RR Lyrae stars in Crater~II and the broad spatial coverage of our observations we can study their radial profile, as a representation of the old population of the galaxy. To do this we took into account the ellipsoidal distribution fitted in the previous section (Table~\ref{tab:sp}) by calculating the elliptical distance ($r'$) as half the geometrical constant of the ellipse at the location of each individual RR Lyrae star.

Following previous studies \citep[e.g.][]{martinez2015}, we use a running average technique to construct the radial profile in order to avoid spurious fluctuations. After sorting the sample of RR Lyrae stars by their $r'$, we counted the number of RR Lyrae stars within a box with a fixed size of $5\arcmin$, which progressively moves in steps of $2\arcmin$ from the center. The profile of the numbers of {\rrab} stars as a function of $r'$ is shown in the top panel of Figure~\ref{fig:Nab_mag_prof}. The shape of the profile does not change significantly with small variations of the chosen step and size of the box. The profile shows several interesting features. First, the RR Lyrae stars are not centrally concentrated. Indeed, there are no RR Lyrae stars within the first $5\arcmin$ from the center of the galaxy. The number of {\rrab } steadily increases from the center until $\sim$20\arcmin, then gradually decays, with a hint of a slight increase near $r'\sim 40\arcmin$. The bottom panel in Figure~\ref{fig:Nab_mag_prof} shows the mean magnitude $\langle i_0 \rangle$ of the {\rrab} stars as a function of $r'$. In this case, the running average was forced to boxes containing 7 stars and the moving step was done every 3 stars. This way we assure that each box had enough stars to get a meaningful mean magnitude. Interestingly, in this profile we note a clear rise of the mean luminosity of the RR Lyrae star population as we move from the center of Crater~II to its outer region. The RR Lyrae stars in the outermost parts of Crater II are, on average, brighter than the inner ones.

In the Sculptor dSph galaxy, \citet{martinez2015} were able to identify different behaviors in the radial profile between the bright/faint (or metal-poor/metal-rich) RR Lyrae stars in that galaxy. This was interpreted as a metallicity (radial) gradient in Sculptor. Following that idea, we split the RR Lyrae stars in Crater II in a bright (B) and faint (F) group. The PLZ in $i$ (equation~\ref{eq:PLi}) at the adopted distance modulus was used to divide the sample, as shown in the top panel of Figure~\ref{fig:BF_profile}. Only {\rrab } were used since types c and d are systematically brighter than the ab type (see Figure~\ref{fig:PL}). The radial profiles of each of these two groups of {\rrab} are displayed in the middle panel of Figure~\ref{fig:BF_profile}. It is clear that the two populations follow different profiles. In particular, the B sample is more extended over the body of Crater~II while the F sample is more centrally concentrated. This can also be seen in the spatial distribution of the two groups (bottom panel of Figure~\ref{fig:BF_profile}) in which is clear that the blue diamonds (B sample) dominate in the outermost parts of Crater II.

There are no clear asymmetries in the spatial distribution of the B and F samples that can be interpreted as the galaxy being spread along the line of sight, which would imply one side of the galaxy  being closer to us than the other. 

Small metallicity differences of just $~\sim 0.17$ dex may account for the observed different profiles. This would imply that the metal poor (brighter) population is more widely distributed than the metal-rich (fainter) population of RR Lyrae stars, which is more centrally concentrated. This behavior has been observed in other galaxies such as the Sculptor dSph \citep{martinez2016a} and Tucana dSph \citep{bernard08}. Evolved RR Lyrae stars or He-enhanced population \citep{marconi18} may be alternative explanations for the existence of the B sample.

The spectroscopic surveys available in Crater II have very limited spatial coverage, with \citet{caldwell17} covering $\sim 1\,r_h$ of the galaxy ($\sim$half the radius covered with DECam) and \citet{fu19} getting spectra only within $15\arcmin$ from the center of Crater II. Nonetheless, \citet{caldwell17} conclude that no radial gradient in metallicity is present in Crater II, based on a sample of 40 red giant stars which are radial velocity members. A more uniform spectroscopic survey in Crater II, covering the outermost parts of the galaxy, is needed to further investigate the presence of a small metallicity gradient as suggested by the distribution of the RR Lyrae stars.

\section{Anomalous Cepheids} \label{sec:AC}

Seven anomalous Cepheids were identified in Crater II (V1, V26, V86, V107, V108, V109 and V110). Anomalous Cepheids are pulsating stars located above the horizontal branch. Although these stars are usually interpreted as tracers of intermediate age population (stars with masses 1-2 $M_\odot$ cross the instability strip during their evolution), they have been found also in purely old stellar population systems \citep[e.g. Sculptor, Sextans, Ursa Minor, Draco, Leo II, see compilation in][]{fiorentino12a}. In these cases, binary evolution is likely the best scenario to explain their origin. A recent discussion on binary evolution channels that may produce Anomalous Cepheids can be found in \citet{gautschy17}. Because Crater II does not appear to contain stellar populations younger than 9 Gyr \citep[see companion paper,][]{walker19}, binary evolution must be responsible for creating the observed anomalous Cepheids in this galaxy.

V1 and V26 were detected by \citet{joo18} but they were tentatively classified as field RR Lyrae stars, although they discussed the possibility that they may indeed be anomalous Cepheids. On the other hand, only V1 is present in the central parts of Crater II. It was measured by \citet{monelli18}, although their classification was set as "Peculiar". V1 is one of the faintest of the anomalous Cepheid stars, and it was argued that it was not bright enough to be of this type. The increased number of stars detected in this part of the CMD in this work leaves no doubt that an important number of anomalous Cepheids are present in Crater II.

The mean $g_0$ magnitude of 5 of the anomalous Cepheids in Crater II range from 19.98 to 20.64, which is 0.3-0.9 mags brighter than the RR Lyrae stars (see Figure~\ref{fig:CMD}). One additional star, V108, is significantly brighter, with $g_0=18.79$, which locates it $\sim 2.1$ mag above the horizontal branch. Anomalous Cepheids are expected to lie $\sim 0.5-2.5$ mag above the horizontal branch \citep{catelan15}, and thus V108 is still within the expected range. Star V1, as well as V26 and V86, have a difference in magnitude $<0.5$ with respect to the horizontal branch. Although this may seem too small as compared with the ranges given in \citet{catelan15}, other stellar systems may have anomalous Cepheid stars in similar locations in the CMD. For example, the Sextans dSph galaxy has 7 anomalous Cepheids, which were recently measured in the $g$ band by \citet{vivas19}, and are located between 0.4 and 1.8 mag above the horizontal branch, similar to the location of the anomalous Cepheids observed in Crater II.

\begin{figure}
\includegraphics[width=1.0\columnwidth]{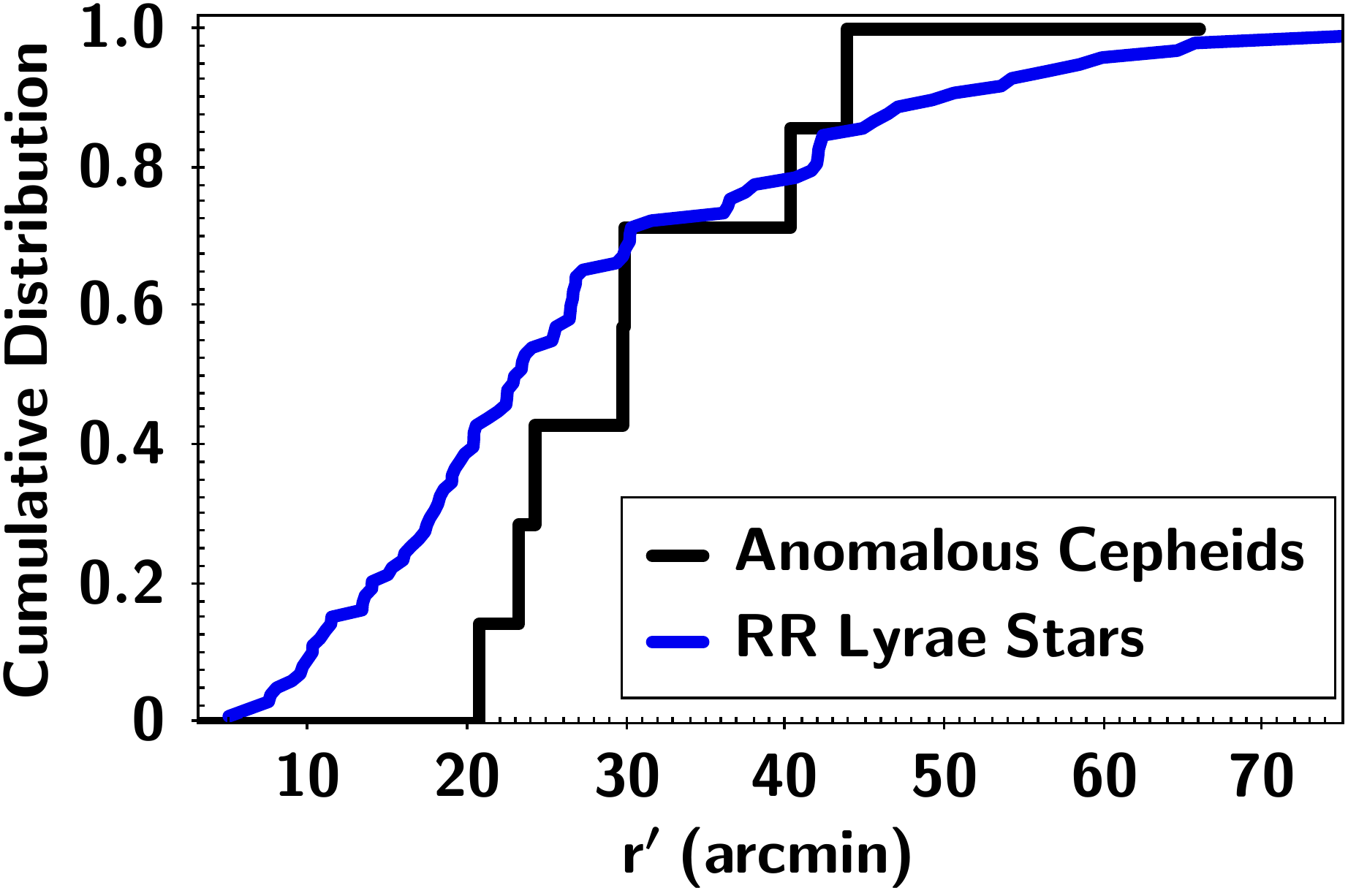}
\caption{Cumulative distribution of RR Lyrae stars and anomalous Cepheids along r', a projection of the ellipsoidal distance to the stars.}
\label{fig:radial}
\end{figure}

In Figure~\ref{fig:radial} we show the cumulative distribution of both RR Lyrae stars and anomalous Cepheids along the
ellipsoidal distance, $r'$. Besides the much smaller number of anomalous Cepheids compared to the RR Lyrae stars, the cumulative ellipsoidal distributions look similar, although anomalous Cepheids are all found within $r'<44\arcmin$ and RR Lyrae stars reach out up to about twice that distance. A Kolmogorov-Smirnov (KS) test indicates that both data samples are indeed drawn from the same distribution. Although it does not constitute a proof on its own, this result is at least consistent with our interpretation above that the anomalous Cepheids should originate from an old population, similar to the RR Lyrae stars. This is not always the case. For example, \citet{fiorentino12a} have shown that the cumulative distributions of anomalous Cepheids and RR Lyrae stars are significantly different in the Large Magellanic Cloud (LMC). It is also different than the distribution of classical Cepheids, suggesting that the origin of anomalous Cepheids in that galaxy cannot be uniquely matched with either a young or old population. Mass segregation could, in principle, add complexity to the comparison between radial distributions since anomalous Cepheids are more massive than RR Lyrae stars. However, this mechanism is not important in a low density galaxy like Crater II in which stellar encounters should be rare. On the other hand, dynamical mass segregation is not expected in dark-matter dominated systems \citep[e.g.,][]{kim15}.

\begin{figure}
\includegraphics[width=0.95\columnwidth]{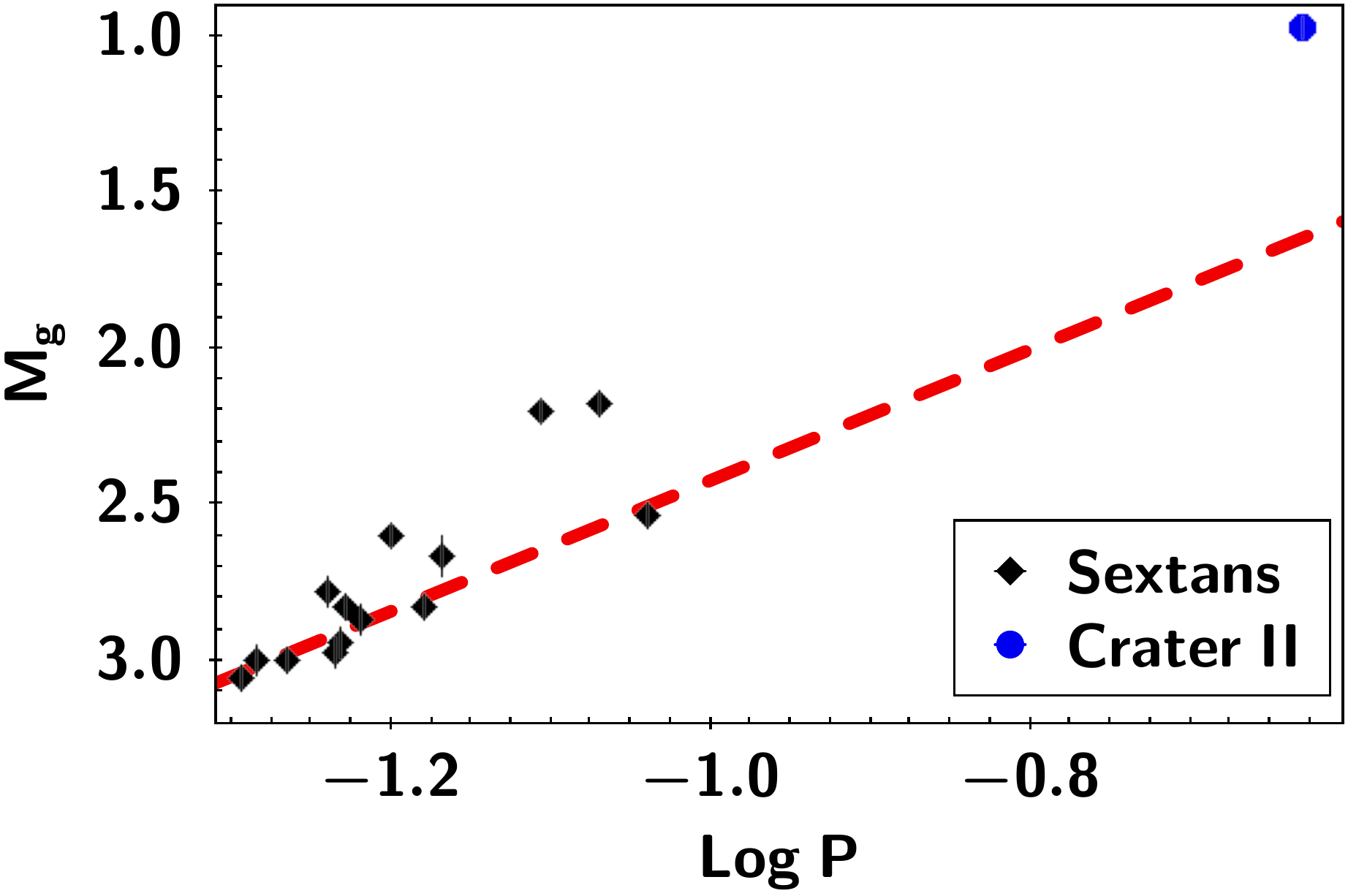}
\caption{PL relationship in the $g$ band for dwarf Cepheid stars in the Sextans dSph galaxy. The red line indicates the best fit obtained for the presumed fundamental mode pulsators. Sextans data and PL relationship are from \citet{vivas19}. The blue circle shows the period and absolute magnitude of the Crater II dwarf Cepheid star, V97. The latter was calculated assuming the distance modulus obtained with the RR Lyrae stars, $\mu_0=20.333$. A shift of $+0.036$ mag was included in the Sextans data and PL relationship to take into account the differences between the Pan-STARRS PS1 system, in which the Sextans data was calibrated, and the SDSS system, in which the Crater II data was calibrated.}
\label{fig:PL_DC}
\end{figure}

\section{Other variables} \label{sec:othervar}

Six dwarf Cepheid stars were found in this work. The main characteristic of these stars is their very short periods. The six stars we found have periods between 0.04 and 0.23 days, and amplitudes in $g$ between 0.15 and 0.74 mag. Only one of them, however, seems to be a member of the Crater II galaxy, This star, V97, has an extinction-corrected mean magnitude of $g_0=21.31$, which locates it at 0.4 mag below the horizontal branch (Figure~\ref{fig:CMD}). Although that may seem as too bright for this type of star, V97 has a rather long period, 0.23 days. Dwarf Cepheids obey a PL relationship, with the brightest stars having longer periods. Thus, it is expected that a star with a period of 0.23d will be one of the most luminous in the class. Unfortunately, there are very limited data available for calibrating the PL relationship of dwarf Cepheids in the DECam passbands. A PL relation in the $g$ band is provided for the dwarf Cepheids in the Sextans dSph galaxy by \citet{vivas19}, but the stars in Sextans span a very limited range in period, from 0.05 to 0.09 days. An extrapolation of that PL to the period of the dwarf Cepheid in Crater II is shown in Figure~\ref{fig:PL_DC}. There is an important discrepancy between V97 and the extrapolation of the PL in Sextans. One possible reason for this is that V97 may not be not pulsating in the fundamental mode, but in the first overtone mode, which will make the PL $\sim0.3-0.5$ mag brighter \citep{nemec94,poleski10}. With the data at hand, it is not possible to know the pulsation mode of V97. Also, it is possible that the slope obtained by \citet{vivas19} for Sextans, $-2.10$, is too shallow because of the limited range of periods available for that galaxy. Indeed, \citet{poleski10} found a slope in the $V$ band of $-2.84$ and $-3.17$ for fundamental and first overtone pulsators in the LMC, which span a range in periods from 0.04-0.25 d. A steeper slope may bring V97 within expectations for dwarf Cepheid stars in Crater II. 

Since the CMD of Crater II does not show signs of a young, metal-rich population, V97 is likely a SX Phoenicis star rather than a $\delta$ Scuti star.

It is worth mentioning that underluminous RR Lyrae stars have been observed. The case of the type c field star TV Lib has been discussed by \citet{bono97}, who suggest that the brightness and period for this star is only compatible if it has a younger age than the bulk of the other field stars. Also, \citet{pietrzynski12} found an underluminous field RR Lyrae in an eclipsing binary system, with a stellar mass that is at least a factor of two smaller than canonical RR Lyrae. This evidence indicates that these variables could also be the progeny of close binary systems that experienced mass exchange for several Gyrs. Although cases like the ones just described are expected to be rare, those channels may be alternative explanations for star V97.

Although the continuous coverage for several hours of our observations make them ideal to identify dwarf Cepheids with periods of only a few hours, our data is not deep enough to uncover the full population of these types of stars in Crater II, if indeed they exist. From \citet{vivas13} and \citet{vivas19}, variables with amplitudes of $>0.2$ mag can be detected if the photometric errors are $\lesssim 0.05$ mag. In our data, that limit occurs at $g\sim 22.0$ (Figure~\ref{fig:error}). On the other hand, dwarf Cepheids can be found down to $\sim 3$ mag below the horizontal branch \citep{vivas13}. Thus, in Crater II there could be dwarf Cepheids as faint as $g\sim 24$, which is beyond our observational limit.

Besides V97, there are other 5 dwarf Cepheid stars in the field, located at different magnitudes above the horizontal branch. These stars seem to be foreground variable stars. Assuming the Sextans PL (which in this case does cover the same period range), those 5 stars should be Halo stars, likely SX Phoenicis stars, located between $\sim8$ to 43 kpc from the Sun.

Finally, 14 eclipsing binaries were also measured in this work, but none of them are located near the main Crater II features in the CMD (Figure~\ref{fig:CMD}), and therefore they are likely foreground Milky Way stars. Among them are several detached, semi-detached, and contact binaries.

\section{Conclusions} \label{sec:discussion}

We present a complete census of periodic variable stars in the ultra diffuse satellite Crater II, covering up to $2\times r_h$. Crater II is very rich in RR Lyrae stars, having 99 stars, of which the majority (84) are of type ab. The low number of type c (5 stars) is likely due to the fact that the horizontal branch of the galaxy ends within the instability strip, in the region of the \rrc. The large number of RR Lyrae stars and the good coverage of the entire galaxy allows us to infer the shape of Crater II, which turns out to be rather elliptical, with an ellipticity of 0.24 and a PA of 153\degr. The period distribution of the {\rrab } is quite narrow and suggests a Oo II classification. However, the distribution of the stars in a Bailey Diagram is more consistent with Oo-intermediate, which is our preferred classification. No HASP stars are present, which is interpreted as this galaxy having an exclusively metal-poor population. Furthermore, by inverting the PL relationship, we were able to infer Crater II has a metallicity dispersion of only 0.17 dex, which is significantly lower than typically found among dSph galaxies.

There is, however, a difference in the spatial distribution of the bright and faint RR Lyrae stars, with the brightest sample dominating in the outermost parts of the galaxy. This suggests multiple populations are present among the oldest stars in the galaxy. Although Crater II has a very small metallicity dispersion, we find that the most metal-poor stars within our sample dominate in the outermost parts of the galaxy. This behavior has been observed in other galaxies and suggests that new stars were born from a more centrally concentrated and slightly enriched medium.

In the companion paper \citep{walker19}, it is shown that Crater II clearly has two sub-giant branches which are well fitted by populations of different ages, 10.5 and 12.5 Gyr, and similar metallicities. The former of those populations, however, may be too young to form RR Lyrae stars. Thus, the RR Lyrae stars discussed in this paper are, in principle, associated exclusively with the older population of Crater II.

Crater II also has an important population of anomalous Cepheids. These stars have been traditionally associated with intermediate-age stellar populations. However, no other clues for such population are present in the CMD of Crater II. These variable stars should have formed exclusively through binary evolution of old stars.

Variable stars, and particularly RR Lyrae stars, are excellent tracers of extra-tidal material around disrupting galaxies. RR Lyrae stars, for example, trace the tidal tails of the Sagittarius dSph galaxy \citep[e.g.,][]{hernitschek17}. Extra-tidal RR Lyrae stars have also been found in other satellites such as Carina \citep{vivas13} and the Hercules UFD \citep{garling18}. Crater II may have been interacting closely with the Milky Way in the past which may have generated tidal tails \citep{fu19}. Since this is an ultra diffuse galaxy, such tidal tails may be hard to detect because of their low surface brightness. Being rich in RR Lyrae stars, a search of extra-tidal material using these stars seems compelling.

\section*{Acknowledgments}

We thank the anonymous referee for a constructive review of the paper and for useful suggestions. 
This project used data obtained with the Dark Energy Camera (DECam),
which was constructed by the Dark Energy Survey (DES) collaboration.
Funding for the DES Projects has been provided by 
the U.S. Department of Energy, 
the U.S. National Science Foundation, 
the Ministry of Science and Education of Spain, 
the Science and Technology Facilities Council of the United Kingdom, 
the Higher Education Funding Council for England, 
the National Center for Supercomputing Applications at the University of Illinois at Urbana-Champaign, 
the Kavli Institute of Cosmological Physics at the University of Chicago, 
the Center for Cosmology and Astro-Particle Physics at the Ohio State University, 
the Mitchell Institute for Fundamental Physics and Astronomy at Texas A\&M University, 
Financiadora de Estudos e Projetos, Funda{\c c}{\~a}o Carlos Chagas Filho de Amparo {\`a} Pesquisa do Estado do Rio de Janeiro, 
Conselho Nacional de Desenvolvimento Cient{\'i}fico e Tecnol{\'o}gico and the Minist{\'e}rio da Ci{\^e}ncia, Tecnologia e Inovac{\~a}o, 
the Deutsche Forschungsgemeinschaft, 
and the Collaborating Institutions in the Dark Energy Survey. 
The Collaborating Institutions are 
Argonne National Laboratory, 
the University of California at Santa Cruz, 
the University of Cambridge, 
Centro de Investigaciones En{\'e}rgeticas, Medioambientales y Tecnol{\'o}gicas-Madrid, 
the University of Chicago, 
University College London, 
the DES-Brazil Consortium, 
the University of Edinburgh, 
the Eidgen{\"o}ssische Technische Hoch\-schule (ETH) Z{\"u}rich, 
Fermi National Accelerator Laboratory, 
the University of Illinois at Urbana-Champaign, 
the Institut de Ci{\`e}ncies de l'Espai (IEEC/CSIC), 
the Institut de F{\'i}sica d'Altes Energies, 
Lawrence Berkeley National Laboratory, 
the Ludwig-Maximilians Universit{\"a}t M{\"u}nchen and the associated Excellence Cluster Universe, 
the University of Michigan, 
{the} National Optical Astronomy Observatory, 
the University of Nottingham, 
the Ohio State University, 
the OzDES Membership Consortium
the University of Pennsylvania, 
the University of Portsmouth, 
SLAC National Accelerator Laboratory, 
Stanford University, 
the University of Sussex, 
and Texas A\&M University.

Based on observations at Cerro Tololo Inter-American Observatory, National Optical
Astronomy Observatory (NOAO Prop. ID: 2017A-0210, PI: Walker), which is operated by the Association of
Universities for Research in Astronomy (AURA) under a cooperative agreement with the
National Science Foundation.

%%%%%%%%%%%%%%%%%%%%%%%%%%%%%%%%%%%%%%%%%%%%%%%%%%

%\bibliographystyle{mnras}
%\bibliography{KV}

%%%%%%%%%%%%%%%%%%%%%%%%%%%%%%%%%%%%%%%%%%%%%%%%%%

\appendix

\section{Notes on Individual variables}
\label{sec:appendix}

V4, V6, V29, V30, V33, V64, V80 were classified by \citet{joo18} as type c, and V34, V82 and V90 were classified as type ab. Our lightcurves for those 10 stars indicate they are pulsating in both modes simultaneously and thus they were re-classified as type d here. Our classification as {\rrd } agrees with the one given by \citet{monelli18} for V29 and V33.
However, V4 was also classified as a type c in \citet{monelli18}.

\citet{joo18} classified V1 and V26 as field RR Lyrae stars. Given their location in the CMD we classified them as anomalous Cepheids. 

V86 was classified by \citet{joo18} as a Crater II \rrc. Although our period and shape of the lightcurve agree with such classification, V86 is $\sim 0.3$ mag brigther (in $g$) than the rest of the stars in the same class. Unfortunately, there is serious overlap between the pulsational properties of anomalous Cepheids and RR Lyrae stars \citep{catelan15}, which is the reason why the former are hard to recognize in the field. As seen in Figure~\ref{fig:CMD}, V86 is the bluemost of the anomalous Cepheids with $(g-i)=0.12$. It is however still within the expected range of the instability strip. The amplitude in $g$ is large (0.77 mag) which is right in the middle of the distribution of amplitudes for anomalous Cepheids in Crater II, which range between 0.45 to 1.12 mag.  {\rrc } stars, on the other hand range between 0.43 to 0.72 mag. Thus, V86 has an amplitude somewhat larger for type c stars, although not completely unusual for these variables, as seen in the Bailey diagrams (Figure~\ref{fig:PA}, particularly the bottom panel). Since the spread in magnitude for the RR Lyrae stars is very tight and V86 is clearly an outlier in the PL relationship for Crater II, we changed its classification to anomalous Cepheid. 

We classified V69, V93 and V94 as a type ab, while \citet{joo18} assigned a type c to them. Similarly, V72, V76 and V92 which were classified as {\rrc } by \citet{monelli18}, are here classified as type ab, in agreement with the classification given by \citet{joo18} for these 3 stars. It is possible that those works detected a 1-day alias of the true period. Our continuous coverage for several hours minimizes the 1-day aliases and give us confidence that our periods are the correct ones.

V97 has an ambiguous classification in \citet{joo18}, either a type ab or a field dwarf Cepheid. The period and the location of the star in the CMD make us believe it is indeed a dwarf Cepheid (a SX Phoenicis) and it is likely part of the Crater II galaxy.

V98 was classified by \citet{monelli18} as a possible periodic star but the noisy light curve did not allow further classification. We were not able to confirm V98 as variable star since it failed to have significant brightness variations simultaneously in both $g$ and $i$ in our data.

V111 is classified here as a field type ab star. However, the Gaia catalog of RR Lyrae stars \citep{clementini19} classifies it as type d with a first overtone period of 0.3855d.

% Don't change these lines
\bsp	% typesetting comment
\label{lastpage}
\end{document}